\documentclass[11pt,twoside,a4paper]{article}
\usepackage{txfonts}
\input{epsf.tex}

\title{Distances to nearby molecular clouds and star forming regions\\
III. Localizing extinction jumps with a Hipparcos calibration of
2mass photometry}
\author{J. Knude\\
Niels Bohr Institute, Copenhagen University\\ 
Juliane Maries Vej 30, DK-2100 Copenhagen {\O}\\
indus@nbi.ku.dk}
\date{June 17, 2010}
\begin{document}
\maketitle
\section{Abstract}
We want to estimate the distance to molecular clouds
in the solar vicinity in a statistically precise way. 
Clouds are recognized as extinction discontinuities. The extinction is 
estimated from the $(H-K) \ vs. \ (J-H)$ diagram and distances from a 
$(J-K)_0 \ vs. \ M_J$ relation based on Hipparcos.
The stellar sample of relevance for the cloud distance is confined by the
FWHM of the  $A_V / D_{\star}(pc)$ or of its derivative. The cloud distance is 
estimated from fitting a function to the $(A_V, 1/ \pi_{JHK})$ pairs in this 
sample with a function like $arctanh^p (D_\star /D_{cloud})$ 
where the power $p$ and $D_{cloud}$ both are estimated. The fit follows the 
$(A_V, 1/\pi_{JHK})_{cloud}$ data rather well. Formal standard deviations less 
than a few times 10 pc seem obtainable implying that cloud distances are 
estimated on the $\lesssim$10$\%$ level. Such a precision allows estimates of 
the depths of cloud complexes in some cases. As examples of our results we 
present distances for $\sim$25 molecular clouds in Table ~\ref{t2}.\\
$Keywords$: interstellar medium: molecular cloud distances

\maketitle

\section{Introduction}

Distances to nearby molecular clouds are essential in many contexts.
The more precisely measured ones are often 
based on dedicated medium band optical photometry of selected stellar types in 
lines of sight in the general direction of the cloud and its immediate 
surroundings. It is an advantage that the optical bands are so sensitive to 
extinction but the same sensitivity of course sets limits on the amount of 
extinction that may be penetrated. 

All photometric systems are not equally suited for extinction purposes 
since a density of sight lines as high as possible is required to decrease 
selection effects and all systems
are not equally useful for classifying all stellar types. The Vilnius system
seems a good choice for optical work because it permits accurate
estimates of intrinsic properties such as absolute magnitude and colors for 
almost any kind of star. The Str{\"{o}}mgren-H$_{\beta}$ system may also be used 
but for a substantially narrower range of spectral types and mainly for main
sequence stars. But it has the great advantage of being based on the extinction
free $\beta$-index.

After the Hipparcos parallaxes, Perryman et al. (\cite{PMAN97}), have become 
available combinations with 
classifications from other sources have been used and resulting in distance $-$ 
extinctions pairs that estimate the distance to the less obscure
parts of molecular clouds, Knude and H{\o}g (1998), Lombardi, Alves and Lada 
(\cite{LAL06}).

Alternatively the parallax data already available may be complemented
with new observations sensitive to the extinction, e.g. polarization as used
by Alves and Franco (\cite{AF06}, ~\cite{AF07}) in investigations of Lupus clouds
and of the Pipe Nebula respectively. Polarization has the advantage
that it may be estimated without any knowledge of the target classification and is
much more precisely measured than photometry. 

A limiting condition of the Hipparcos parallaxes is that they pertain to
fairly bright stars measured in the optical and consequently are confined to the
low extinction parts of the clouds and only may be used for clouds in
the immediate solar vicinity.

If a distance estimate to a cloud is requested photometry is required for a
substantial 
number of stars. Such observations may be rather time consuming despite the
advantages brought about by CCD photometry. It would therefore be convenient if a 
method exploiting available all sky photometric data might be established. It 
only requires
that the photometry may be dereddened and that the dereddened colors may be 
calibrated in terms of absolute magnitude.  

Near infrared data may not be the obvious choice for extinction estimates but 
some sensitivity to reddening is left and one benefits from 
the much better penetrating power of the NIR data so the association of the 
data to the molecular cloud is possibly better established than that of the 
optical data. Infrared 
data have been widely used to produce the projection of extinction on the sky 
in the form of impressive maps and less used for distance determination,
e.g. Lombardi, Lada and Alves (\cite{LLA08}).
  
For each starget, distance and intrinsic colors should result somehow and the
combination of many sight lines may provide a statistical estimate of the cloud distance. As 
we will notice in the following discussion several regions known to contain 
molecular material do show an extinction discontinuity at some not very precise
distance. The cloud 
distance may be estimated by the eye but we have investigated some quantitative
statistical methods from which the distance intuitively may be estimated --
but these
methods do not always work in a satisfactory way. Even by limiting the study to
the most accurate data, we can not be sure that the data are statistically
significant and statistics as the mean, median, standard deviation, 
$\sigma_{A_V}$/$A_V$ versus distance may have shortcomings so they do not
immediately guarantee a representative observation of the dust distribution and
in particular they do not provide an estimate of the uncertainty of the suggested
distance. To meet the required error assessment we suggest instead that some 
analytical function is fitted to the data defining the extinction jump and that
the error may be estimated by the standard deviation from the distance fit.
We propose that the sample pertinent for a distance derivation may be extracted
from the variation of the line of sight density in a consequential manner. If 
all stars used to define the jump really are located at a well defined distance, 
the uncertainty of the estimated cloud distance is on the $\pm$10$\%$ level or 
better. Due to selection effects, some of which are 
introduced by limiting the $JHK$ photometry to $\sigma_{JHK} \leq 0.040$ or 
$\sigma_{JHK} \leq 0.080$, originating in the way the sample used for fitting the
variation of extinction with distance is defined, the distance estimate may not 
be robust. But we think that the way we extract this sample -- from the
variation of the average line of sight density with distance -- may be a good 
approximation to a robust method. At least it is systematic and not based on 
any personal judgement. Biases are introduced by the co-incidence of 
the main sequence and giant relations in part of the $(H-K)_0 - (J-H)_0$a
diagram. The absence of some stellar classes, e.g. 
$\sim$G6 -- M0, with a certain range of absolute magnitudes, causes the rise of 
extinction with distance to be more shallow than
expected when a molecular cloud is encountered. This have consequences for the 
statistics and for the estimated cloud distance but we suggest a way to include
these stars after the variation of extinction with distance has been computed 
from the stars earlier than $\sim$G6 and later than $\sim$M4.        

The paper falls into four parts: 
In Section~\ref{How} we consider ways and means to estimate 
$D_{cloud} \pm \sigma_{D_{cloud}}$. We present a discussion of about 25 
cloud distances in Section ~\ref{25clouds} and the resulting distances are 
summarized in Section ~\ref{summarizing}, Table~\ref{t2}.

We have banished the gory details of the main sequence calibration to 
Appendix ~\ref{appA}. The discussion of which 2mass stars that may be used 
for estimating $A_J$ and $M_J$ is deferred to Appendix ~\ref{appB}.

\section{How to estimate the cloud distance? Serpens region as a template}\label{How}

In the following we consider various ways a cloud distance may be estimated and 
present a procedure we suggest to use with the calibrated 2mass 
photometry. For details pls. refer to Appendix ~\ref{appB}.
 
\subsection{Cloud distance estimate from A$_{\bf V}$(mean), A$_{\bf V}$(median), 
$\bf \sigma_{A_V}$ vs.  distance}
We confine the sample by the photometric precision, quality as well as
multiplicity flags and start by including lines of sight
outside the frayed cloud confinement. The sample with counts less than the
average count, 100/reseau by definition, minus one $\sigma_{count}$ may indicate
cloud directions to a better degree but here we show the result independently of 
the reseau counts. Including all lines of sight is normally not justified but in the
case of Serpens A and B where the preselected solid angles match the clouds well
it seems acceptable, see Fig.~\ref{f16} displaying the distribution of counts.
The resulting distances and extinctions are in Fig.~\ref{f18}, \ref{f19} 
indicating a steep rise to $A_V\approx$2.5 at distances between 160 and 200 pc. 

Only 2mass data better than 
$\sigma_{JHK} < $ 0.040 has been used. The eye will probably estimate the cloud 
distance to be somewhere between 150 and 200 pc. Strai$\check{z}$ys et al. 
(\cite{straizys96}) measure a distance 260$\pm$10 pc to this region. The 
diagram, Fig.~\ref{f19}, shows a few auxiliary curves. The two dashed curves indicate
the maximum measurable extinction for the values 11.0 and 14.6 for $V-M_V+5$ 
that may be traced by a M4V and a M0V star respectively in a sample with 
$J_{lim}$ = 14.5 mag. We see that the 
late M4 -- T dwarfs are well confined by such a maximum extinction curve.
We also note that the group have a well defined minimum extinction in the distance range from 200 to 400 pc at which distance the minimum extinction starts
rising. The extinction discontinuity is well defined by the data. The early and 
late groups suggest that extinctions between $\approx$0 and $\approx$2.5 mag
are present in the distance range from $\approx$60 to $\approx$400 pc. Within
this box the potential K dwarfs are extracted and the Figure shows that these
K dwarf candidates support the presence and location of the extinction 
discontinuity. The median and mean the extinction are shown, computed for 20 
pc bins and in 10 pc steps. Beyond about 400 pc the two values stay identical.
$\sigma_{A_V}/\sqrt{N-1}$ where $\sigma_{A_V}$ is the standard deviation and N 
the number of stars in the distance bin is also indicated. $\sigma_{A_V}$ is 
computed in the same intervals as the mean and median. For this field the 
error of the mean seems to follow the rise of the median extinction. 
One might think of using some combination of 
$\sigma_{A_V}$ and $A_V$ vs. distance to signify the onset of molecular 
extinction (Padoan, Nordlund, Jones (\cite{PNJ97}) and the 
Lada et al. (\cite{LLCB94}) $\sigma_{A_V}$ vs. $A_V$ variation). Between 600 and 
1200 pc the median has a constant slope implying a constant dust density beyond 
the Serpens Cloud and with a known gas/dust ratio the average line
of sight number density of hydrogen may be determined. See the discussion in 
the next Section on how the variation of the line of sight mean density may 
be used to locate the cloud. 

\begin{figure}
\epsfxsize=8.0cm
\epsfysize=16.0cm
\epsfbox{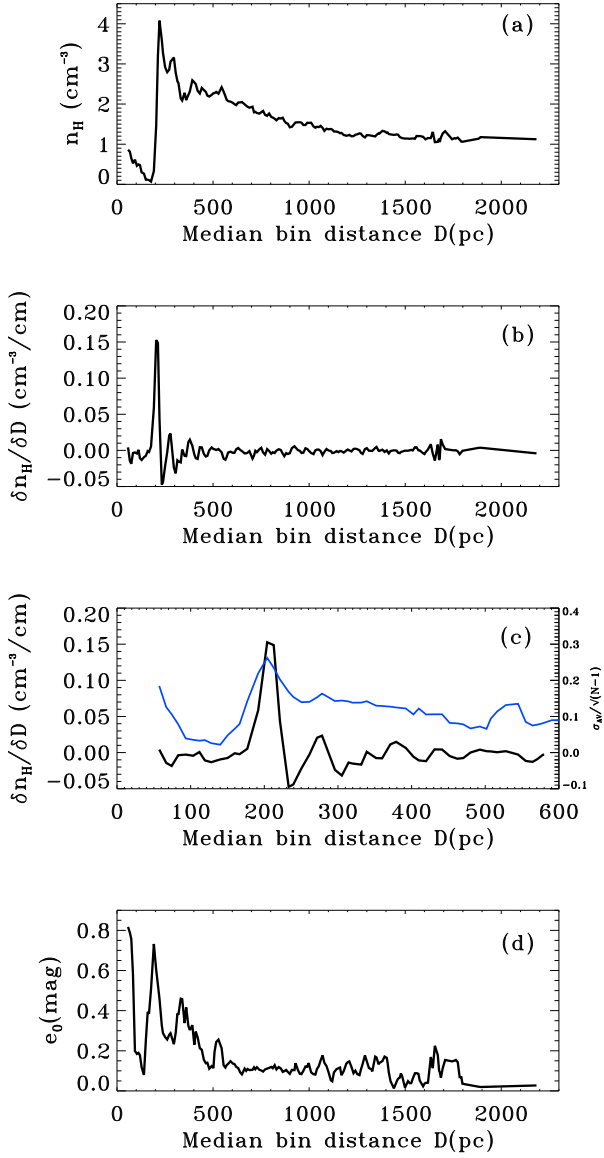} 
\caption[]{Statistics for the Serpens 2by2 region. {\bf(a)} Mean line of sight density 
$n_H$ $cm^\mathrm{-3}$.
{\bf (b)} The derivative of the mean density. {\bf (c)} Zoom of {\bf (b)} and
$\sigma_{A_V}/\sqrt{N-1}$ is plotted as the upper curve and is to the right 
hand scale. {\bf (d)} Mean V-extinction, $e_\mathrm{0}$ in a standard cloud} 
\label{f20}
\end{figure}

The Serpens 2by2 region may be particularly well behaved since the both the mean
and median $A_V$ starts rising at 200 pc as do $\sigma_{A_V}/\sqrt{N-1}$.
It is, however, difficult to quantify the cloud distance and its uncertainty 
from e.g. the median extinction's variation. An average of the distances where
the median starts and stops rising could possibly be used as the distance and
half their difference as an indication of the uncertainty.
As Fig.~\ref{f19} indicates the distance to the cloud is based on all three groups of 
stars implying that the relative error on the individual stellar
distances formally range from $\approx$10$\%$ to $\approx$40$\%$ with an overweight of the
smaller ones.

\subsection{Distance indications from other statistics}

When approaching a molecular cloud the interstellar density
will jump up when the cloud is penetrated. When the density increase is large 
enough over a short distance the increase is reflected in a rise in $A_V$
despite $A_V$ is the integrated of the density along the line of sight.
In order to cause a discontinuity the cloud extinction, sampled over a short
distance, must be comparable to or exceed the extinction accumulated along the 
line of sight to the cloud. With enough data to form 
derivatives we would expect the derivative $\partial n_H / \partial D$ where
$n_H$ $cm^\mathrm{-3}$ is the average number density of atomic hydrogen and $D$ 
the distance to show a dramatic increase over a short distance range.
$n_H$ $cm^\mathrm{-3}$ is formed by converting $A_V$ to $N_H$ with the 
canonical gas to dust ratio. Fig.~\ref{f20}(a) shows the variation of the line 
of sight density of neutral hydrogen for the Serpens 2by2 region and a very 
sharp increase is noticed at $\sim$200 pc. The asymptotic value of $n_H$ is
$\approx$1 atom/cm$^{\rm -3}$ fairly close the the mean density of the diffuse
interstellar medium in the solar vicinity. The constant part of the tail results
when the clouds contribution to the average line of sight density becomes 
negligibel compared to the contribution from the intercloud diffuse medium: 
$N_H(cloud)/D < n_H(intercloud)$. Identical to the distance range where the 
slope of $A_V(median)$ becomes vitually constant. The (b) part of the same 
figure is the derivative of the density with respect to distance (in cm) and 
again we notice a change at $\sim$200 pc. Part (c) of Fig.~\ref{f20} is a zoom 
of the (b) frame displaying the effect of an increasing density over the 
distance range from $\sim$180 to $\sim$210 pc. The (c)
frame also contains a sort of mean dispersion $\sigma_{A_V}/\sqrt{N-1}$ 
pertaing to the 20 pc distance bins.  

Early studies of the patchy structure of the ISM, assuming that the
interstellar medium was constituted by 
a single (or two) type(s) of interstellar cloud(s) floating in an intercloud medium
provided a detailed statistical method to estimate the extinction $e_0$ in the
characteristic cloud, M{\"{u}}nch (\cite{GM52}). This method requires a data set 
of distance and extinction pairs, just what we get from the present study. The
characteristic cloud extinction is given by the expression
$e_0 = M_2/M_1-M_1(1+(\Delta D/D)^2/12)$. $M_1$ is average $A_V$ and $M_2$ 
equals the average of $A_V^2$ in a distance interval $\Delta D$ wide and 
centered on the distance D.
Frame (d) of Fig.~\ref{f20} displays this simple statistics. The
$e_0$ expression is valid when $(\Delta D/D)^2/12)$ is less than unity which is not 
quite the case for the first distance bins. In these bins $\sigma_{A_V}$ is large 
which together with the small average extinctions raises the $e_0$ estimates.
Beyond $\sim$500 pc $e_0$ becomes constant, settling around $A_V$=0.1 mag. 
Converting to a color excess in the $uvby\beta$ system the characteristic cloud reddening
becomes $\approx$0.040, the exact value depends on the choice of $R_V$.
The $e_0 \approx$0.04 is close to the values ranging from 0.025 to 0.045 
calculated from $uvby\beta$ photometry of F stars within 150 pc, 
Knude (\cite{knude79c}). Measured reddenings of \ ~"isolated" clouds were in 
the range from $\sim0.02$ to $\sim$0.11 , Knude (\cite{knude79a}). This 
coincidence is taken as evidence that in a statistical sense our present 
extinction and distance estimates imply results comparable to those obtained 
by independent methods.  

Frame (d) of Fig.~\ref{f20} further contains three peaks at 195, 335 and 530 pc
respectively which probably may be taken as evidence for the presence of molecular
clouds, at least for the 195 and 335 peak's part. That the large $e_0$ values popping
up in a few adjacent distance bins may indicate the distance to a molecular cloud
may not be unexpected after all. When $(\Delta D/D)^2/12)$ $<<$ 1 $e_0 
\approx M_2/M_1-M_1 = \sigma_{A_V}^2/A_V = \sigma_{A_V} \times \sigma_{A_V}/A_V$.  
And according to Fig.~\ref{f19}  $\sigma_{A_V}$ as well as $\sigma_{A_V}/A_V$ have local
maxima at $\approx$200 pc. The 335 pc peak may be an artefact caused by a local 
minimum in the median $A_V$ and there is no local maxima in $\sigma_{A_V}/A_V$ at
this distance. The $\sigma_{A_V}/A_V$ minimum is possibly not real since stars at
335 pc with $A_V$ = 3 -- 4 is not measurable by our method: the missing M0 -- M4 
dwarfs (see the discussion of Fig.~\ref{f18}).  

A well behaved discontinuity as the one in the 2by2 Serpens region
offers several options for the distance estimate: mean and median of
$A_V$, $\sigma_{A_V}$, the mean dispersion $\sigma_{A_V} / \sqrt{N-1}$, 
the mean line of sight density $n_H$, $\partial n_H / \partial D$, $e_0$ or 
equivalently $\sigma_{A_V} \times \sigma_{A_V}/A_V$. Of these $n_H$ and 
$\partial n_H / \partial D$ display sharp peaks at what we interpret as
the cloud distance. The mean dispersion of $A_V$ have a broader peak than 
the derivative of the mean line of sight density. These estimators do 
not provide an immediate uncertainty on the distance but indicate a distance 
range in which the cloud is located. 

\subsection{An algoritm fitting the extinction -- distance variation at an 
extinction jump}\label{AnAlgo}

Due to the rather few distance -- extinction pairs that most often have been
available in the direction of a cloud most studies of cloud distances 
suggest that the cloud distance may be estimated from the distance where the 
increased extinction is first noticed and the location of this rise is
furthermore estimated by the eye. This would of course be correct 
if the stellar distances were perfect with only negligibel errors. 
Other studies claim to have a stellar density high enough to identify 
the backside of the extinction rise as well as the front, Whittet et al.
(\cite{W97}) for the Chamaeleon II cloud, and equivalates the cloud distance to 
the mean of these two distances thus also implying an uncertainty of the
distance estimate.  

With the 2mass data we may often have an observed stellar density that is
higher than otherwise have been the case and we may consider a more 
quantitative approach. 

The $A_V$ vs. stellar distance diagram
is characterized by a set of stars in front of the cloud
measuring only the extinction of the diffuse interstellar medium until the
cloud is reached when the extinction diplays a steep rise over a short
distance range.

The extinction rise shall be matched by
a function staying $\sim$constant until it displays an almost vertical growth.
A horizontal and a vertical line have been used to match thess trends but 
in particular the vertical part seems difficult to accomodate in a systematic 
and robust way. A critical issue is how far beyond the rise stars
can be included in the distance determination?

A function $arctanh^p(x)$ with $0 \leq x < 1$ simulates a combination of a 
horizontal and a vertical line rather well.
And yes, there may be other functions serving our purposes. Our choice is not
completely arbitary as judged from the standar deviations obtained.
Its logarithmic presentation $arctanh(x)=0.5 \times log_e \frac{1+x}{1-x}$ where 
$x$ is short for $\frac{stellar ~distance}{cloud ~distance}$ and $p$ is a power
introduced to emphasize either the vertical trend or the horizontal one 
whichever the least square procedure selects. In order to use the logarithmic
expression we must introduce a maximum distance beyond which no stars are 
included in order to keep the parameter $x = D_{\star}/D_{max}$ less than
unity. NOTE: $D_{max}$ is not the cloud distance but defines the sample used to
estimate $D_{cloud} \ < D_{max}$.

\begin{figure}
\epsfxsize=8.0cm
\epsfysize=8.0cm
\epsfbox{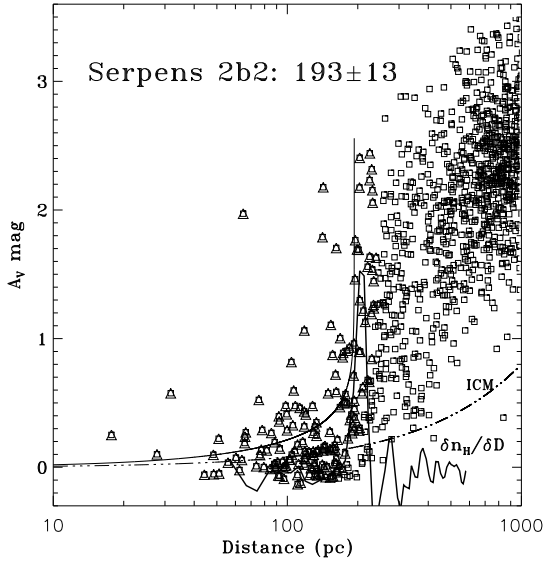} 
\caption[]{The Serpens 2by2 region with the fitted
$arctanh^p(\frac{D_\star}{D_{cld}})$ function shown as the solid curve.
The sample used for the fit is overplotted with triangles. For reference a 
scaled version of $\partial n_H / \partial D$ is shown. The dash -- dot curve 
illustrates a previous estimate of the intercloud density, Knude (\cite{K79}).
The Serpens 2by2 template data are also treated in App.~\ref{appB} Fig.~\ref{f19}} \label{f21}
\end{figure}

A non trivial problem is, however, to define the sample to be included in
the fitting procedure. 
It is a question of how large distances can be included and still be 
pertinent for the cloud distance. Stars far behind the cloud have the large cloud
extinction plus a contribution from the diffuse medium but should not 
enter the cloud distance determination. As seen in Fig.~\ref{f19} the jump contains
several G6V-M0V stars that have a calibration standard deviation of only 
$\sigma_{M_J} \approx$0.1 mag equivalent to a distance uncertainty $\approx$10$\%$,
see Fig.~\ref{f8}. If the
Serpens 2by2 cloud is at 200 pc we should include stars in the interval from 180
to 220 pc. In order to exclude less reddened stars probably beyond the cloud
distance and not including distant stars showing the extinctions in the jump but
not assisting assessing the cloud distance we make a selection 
for the fit.  For the selection we use a curve $A_V ~vs. 
~arctanh^p(\frac{D_\star}{D_{max}})$ to set an upper
distance for each $A_V$. After some experimenting our choice is $p$=4 since this 
value emphasize the shallow part of the data. Note that p=4 is only used for 
selecting the cloud sample when the cloud distance is derived from the curve fitting
p$\pm \sigma_p$ also results. From the density variation in the Serpens region 
$D_{max}$ becomes 250. The 250 pc is not a general upper limit for stars included 
in the curve fitting: In Fig.~\ref{f21} we notice that the requirement 
$A_{V}(\star) > arctanh^4(\frac{D_\star}{D_{max=250}})$ excludes several
stars $D_* \lesssim 200$ pc with a low extinction. 

A systematic definition of the fitting sample is required and should be 
independent of any personal judgement. $D_{max}$ is
determined from the variation of the line of sight average density $n_H$ or 
its derivative $vs.$ distance and is formally defined as the maximum of the 
FWHM points. For Serpens $\sim$250 pc is the largest of the FWHM distances. 
We confine the fitting sample to the stars that are closer and 
more extincted than indicated by the curve 
$A_V(sample \ confinement)$ = $arctanh^4(\frac{D_\star}{D_{FWHM,max}=250})$.

\vspace{0.5cm}
{\it A procedure proposed to estimate the cloud distance:

$\bullet$ confine the cloud on the sky:  contours from star counts or the average
of the $(H-K)$ color formed in reseaus. $\overline{H-K}_{res}$ is preferred to star 
counts since it appears to be more directly linked to the extinction.
The reseau is dynamically defined
to have a radius implying 100 stars/reseau on the average. The minimum cloud 
$\overline{H-K}_{res}$ is estimated from $\overline{H-K}_{res}$ vs. position
diagrams as the $\approx$maximum
of the almost constant value of $\overline{H-K}_{res}$ outside the cloud. 
Fig.~\ref{f28} is an example where $\overline{H-K}_{res}$ = 0.23 is evident as the
maximum for lines of
sight $b \lesssim$ -12$^{\circ}$. All lines of sight with a reseau average exceeding this
$\approx$maximum are accepted as pertaining to the cloud.  

$\bullet$ run codes on the contour sample extracting stars from the $H-K \ vs. 
\ J-H$ diagram: $O - G6$ (primary), $M4 - T$ (secondary), $G6 - M0$ (tertiary) to
estimate $(J-H)_0$ and $M_J$ and compute the $(D_{\star}, A_V(\star))$ pairs.

$\bullet$ bin distance range and use $(D_{\star}, A_V(\star))$ to compute
$(D_{\star}(median), \overline{n}_H (los, median))$

$\bullet$ see if \ $\overline{n}_H (los, median)$ or $\delta \overline{n}_H /
\delta D_{\star}(median)$ displays a peak. Use $D_{FWHM,max}$ of the 
$\overline{n}_H$ or $\delta \overline{n}_H / \delta D_{\star}(median)$ 
variation with distance to confine the sample to be used for the curve fitting: 
$A_{V}(\star) > arctanh^4(\frac{D_\star}{D_{FWHM,max}})$. Note: in the selection
p=4. 

$\overline{n}_H$
is proportional to $A_V(\star) / D_{\star}$. $\overline{n}_H$ is
computed from the median $A_V$ in distance bins but if the density of data points
is not sufficient for the $\overline{n}_H$ vs. $D_{median}$ variation
is replaced by the distribution of individual values $(A_V / D)_{\star}$ from
which $\sim D_{FWHM,max}$ is estimated.

$\bullet$ fit $A_V(jump)$ = $arctanh^p(\frac{D_\star}{D_{cloud}})$ to this
sample. $D_{cloud} \pm \sigma_{D_{cloud}}$ and $p \pm \sigma_p$is returned.
The procedure used for fitting $D_{cloud}$ and $p$ is an implementation of the 
nonlinear Marquardt-Levenberg algoritm. The algorithm varies the parameters 
$D_{cloud}$ and $p$ in search of the minimum in the sum of the squared residuals. 
The iteration stops when convergence is attained.  

$\bullet$ the contour $\overline{H-K}_{res, \ CLOUD}$ defining the cloud perimeter
and $D_{max}$ are the two most critical parameters and must be estimated with care}

\vspace{0.5cm}
As a template we use the Serpens data. The derivative's,
$\partial n_H / \partial D$, variation with distance implies $D_{max}$=250 pc.
The sample are stars above the curve $A_V = ~arctanh^4(\frac{D_\star}{250})$.
The fitting sample is shown in Fig.~\ref{f21} as the combined 
squares/triangles. Then the fitting procedure is run. The convergence normally 
takes place after $\lesssim$10 iterations. For the Serpens 2by2 area the final 
distance becomes $D_{cloud}$=193$\pm$13 pc. The resulting fit is shown 
in Fig.~\ref{f21} together with $\partial n_H / \partial D$, there
is a good coincidence between the resulting cloud distance and the
location of the $(\partial n_H / \partial D)$ peak: the 
$(\partial n_H / \partial D)$ peak may be used to indicate the approximate
cloud location and not least provides the distance range in which the cloud is
situated.

\section{Examples of cloud distances}\label{25clouds}

Considering the multitude of local interstellar clouds we have to be 
selective and only consider a few examples, $\sim$25, of cloud distance determination
from the $JHK$ 2mass data. Most interesting among the local clouds are the
star forming molecular clouds since many of the parameters needed for 
understanding the importance of the environment for the onset of the star 
formation depend on some power of the distance. The distance is of course also 
an important issue when model parameters e.g. from evolutionary 
models of proto and PMS stars are to be compared with observational data.  

During the work with the proposed method a new opportunity for checking the suggested
distances has become available with the advent of the VLBI/VLA astrometric
observations of PMS stars and masers resulting in unprecedented parallaxes to targets
probably associated to star forming clouds.   

\subsection{The Taurus star forming region}

The region covers a substantial part of the sky with longitude ranging from
$\sim$154$^{\circ}$ to $\sim$180$^{\circ}$ and latitude from -24 to the galactic plane. 
We have distributed 16 2$\times$2 $\Box^{\circ}$ regions cowering the main features of
the cloud as indicated by CO intensity maps and compiled 2mass data with
$\sigma_{JHK}$ better than 0.080 mag. Taurus is of special interest since 
its distance has been measured to 137 pc using VLBA astrometry resulting in $\sim$one
percentage accuracy which is an order of magnitude better than what has been 
obtained previously, Torres et al. (\cite{TLMR07}). The 
VLBA astrometry tracking the path on the sky resulting from the yearly 
and proper motion of naked T Tauri stars in the cloud provides individual 
distances with a precision better than 1 pc. Furthermore such a precision allows
the depth of the cloud to be no less than about 20 pc. A mean distance of 137 pc
corroborates the 139$\pm$10 pc deduced from approximate parallaxes based on 
proper motions, Bertout and Genova (\cite{BG06}).  

\begin{figure}
\epsfxsize=8.0cm
\epsfysize=20.0cm
\epsfbox{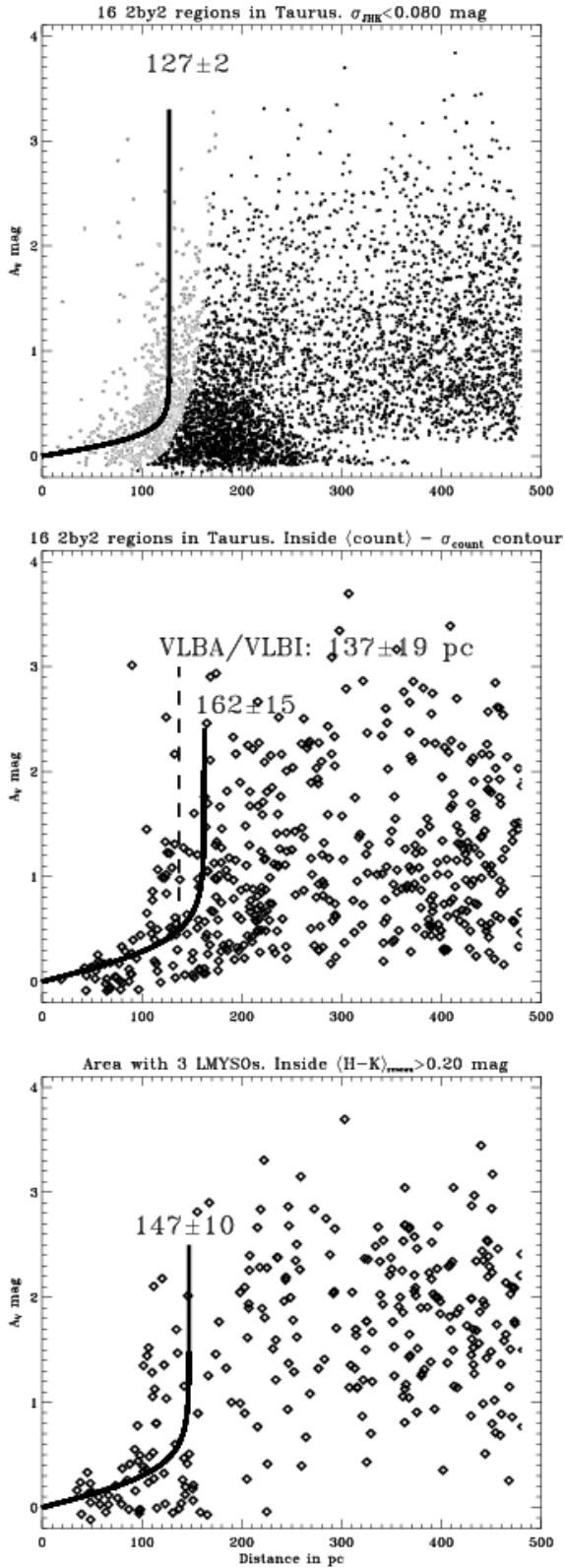} 
\caption[]{Three different selections of a Taurus sample. The $lower \ panel$ 
displays the outcome of the method we suggest to use. $Upper ~panel$: Stellar 
sample with $\sigma_{JHK}\leq$0.080 mag distributed in 16 
2$\times$2 $\Box^{\circ}$ in Taurus. Cloud fitting sample are the light gray
points. $Middle ~panel$: stars in the 16 2$\times$2 $\Box^{\circ}$ in Taurus 
within the ($<$count$>_{res}$ - 1$\times \sigma_{count}$) contour but including 
$\sigma_{JHK}\leq$0.080. $Lower ~panel$: Region containing the three low mass 
YSOs HP Tau, Hubble 4, HDE 283572 for which VLBA astrometry has been performed. 
See Fig.~2 of Loinard et al. (\cite{LTMR08b}). The 137$\pm$19 pc indicated in 
the $middle ~panel$ is from Loinard et al. and the $\pm$19 pc is meant to i
ndicate the possible depth of the complex of clouds. The $lower \ panel$ data 
ar for longitude range [167$^{\circ}$, 177$^{\circ}$] and latitude range 
[-16.8,-14.8], the region containing the T Tau stars with 
VLBA/VLI parallaxes. Only stars with $\sigma_{JHK}\leq$0.040 mag and located 
in reseaus with $\overline{H-K}_{reseau}$ $>$ 0.20. With $D_{max}$ = 250 pc 
the resulting fit is $D_{Taurus}$ = 147$\pm$10 pc which we propose for the 
Taurus distance} \label{f22}
\end{figure}

Confining the sample by the photometric precision alone may include lines of sight
outside the frayed cloud confinement. The sample with counts less than the 
average count, 100/reseau by definition, minus one $\sigma_{count}$ may indicate
proper cloud directions to a better degree. The outcome is shown in the middle 
panel of Fig.~\ref{f22} indicating a steep rise to $A_V\approx$1.5 at distances 
between 100 and 120 pc. Extinctions exceeding 2 mag are noticed for the same 
distance range. Another steep rise is noticed at $\approx$170 pc increasing $A_V$ 
from $\sim$1.5 to $\sim$3.0. We can
not decide whether this dual structure is due to the distribution in depth of 
the Taurus complex or is an effect of the incompleteness of the tracing sample 
causing the sloping appearance of the extinction variation with distance as
discussed in Fig.~\ref{f18}. 

The upper panel of Fig.~\ref{f22} is the extinction variation from a combination of all
data in the 64 square degrees from the 16 2$\times$2 $\Box^{\circ}$ areas without
considering the cloud containment neither from the reseau counts nor from a lower 
$\overline{(H-K)}_{reseau}$ limit. A rather well defined peak in the average lines 
of sight density has an upper FWFM distance at $D_{max} \ \approx$ 200 pc. The 
resulting fitting sample is marked as gray points in the
upper panel and from the curve fit a distance $D_{Taurus}$=127$\pm$2 pc is 
computed. The small dispersion is caused by the large number of stars in the 
sample. Notice that a substantial number of nearby low extinction 
stars are not included in the curve
fitting. Also notice a number of stars at $\approx$80 pc with extinction larger
than 1 mag. These small distances displaying large extinctions are possibly
due to giants mistaken for dwarfs. 

In the central panel the sample is constrained to the stars with reseaus with
counts less than ($<$count$>$ - 1$\times \sigma_{< count >}$). $D_{max}$ has now 
increased to $\approx$300 pc and $D_{Taurus}$=162$\pm$15 pc is computed. The 
vertical dashed line at 137 pc is the average of the VLBA/VLBI paralaxes and the
dispersion $\pm$19 pc is an indication of the depth of the Taurus complex from 
these precise data.

The lower panel is perhaps the most interesting one since it covers the region where
the three low mass YSOs with VLBA/VLBI parallaxes are located. The data are now 
confined by two criteria: $\sigma_{JHK}\leq$0.040 mag and 
$\overline{(H-K)}_{reseau}$ $>$ 0.20. $D_{max}$ = 250 pc, which comply to the
formal definition of $D_{max}$, implies $D_{Taurus}$ = 147$\pm$10 pc. Reducing 
$D_{max}$ with with 25 and 50 pc changes the $D_{Taurus}$ estimate to 130 and 125 pc
respectivly without changing the standard deviation. We suggest 147$\pm$10 pc as
representative and this distance is furthermore only one sigma separated from the
VLBA distance of 137 pc.   

The three different ways of selecting the data from which the fitting sample was
selected result in three different distance estimates for Taurus: 127$\pm$2, 
162$\pm$15 and 147$\pm$10 pc and illustrate the importance of being systematic. The 
distance resulting from our procedure 147$\pm$10 pc is fortunately the one
agreeing best to the VLBA/VLBI parallax 137 pc.
   
\subsection{The $\rho$ Ophiuchus star forming region}

\begin{figure}
\epsfxsize=8.0cm
\epsfysize=8.0cm
\epsfbox{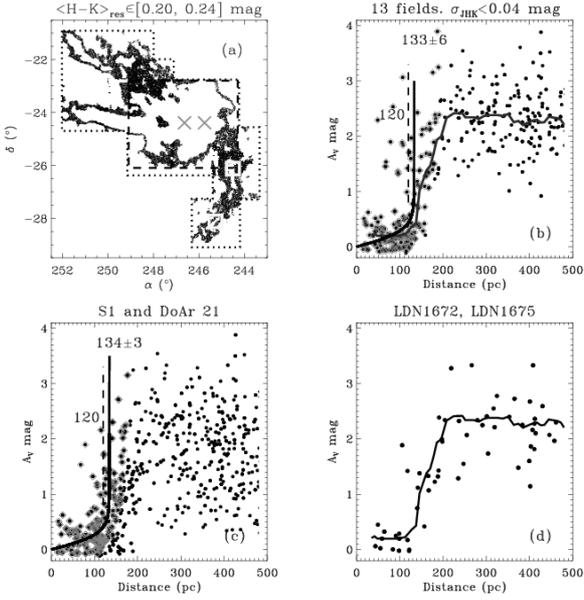} 
\caption[]{Ophiuchus. (a) Dotted boxes are the total area searched in 2mass. 
The perimeter is made up by the stars with $\sigma_{JHK} \ <$ 0.080 mag and 
$\overline{H-K}_{res} \ \in$ [ 0.20, 0.24]. The central dashed box confines 
the $\sim$1.5 $\Box ^{\circ}$ covering L1688, the dense core of the $\rho$ Oph 
complex and contains Oph~S1 and DoAr~21 marked by the two crosses.
(b) Extinction - distance pairs for stars in reseaus with 
$\overline{H-K}_{res} >$ 0.20 mag and with  $\sigma_{JHK} <$ 0.040 mag. The 
solid curve indicates the variation of the median extinction. The sample used 
for the curve fitting is inscribed in diamonds. We note that the fitted 
distance corresponds to the distance where the median extinction starts 
rising. (c) Same as for panel (b) but without any restrictions on 
$\overline{H-K}_{res}$. The dashed line at 120 pc indicates the lower limit
of the range 120 -- 150 pc suggested by Knude and H{\o}g (\cite{knude98}). 
(d) Extinction vs. distance for the two south eastern boxes covering the Arc. 
Mainly the clouds L1675 and L1672 (upper and lower square in (a) respectively)} \label{f25}
\end{figure}

Similarly we have used the extinction map by Cambr{\'{e}}sy (\cite{C99}) 
for the $\rho$ Ophiuchus complex of clouds to define the solid
angle confining the extinction associated with $\rho$ Oph. In this area
we have extracted the 2mass data with $\sigma_{JHK} \leq$0.080 mag. 
We may refine this sample by changing the area, error and value of 
the reseau means. Fig.~\ref{f25}(a) shows the combined area for which data are
extracted (dotted outline), 
The area covering the core region, LDN 1688, and containing the two low mass
YSOs, whose
positions are marked by crosses, is shown as the dashed confinement. The two
squares in the southern extention indicate LDN 1672 (the southern
one) and LDN 1675 respectively. According to Cambr{\'{e}}sy's map the extinction
through the southern feature does not reach the blocking extinction met in the 
cloud core and may therefor suit our approach better -- that is if all the 
clouds are spatially associated. 
Fig.~\ref{f25}(b) shows the resulting distance -- extinction diagram. After confining
the sample to the most precise photometry, $\sigma_{JHK} <$ 0.040 mag and only
using stars in reseaus where the reseau mean exceeds 0.20 mag. The outlines in
Fig.~\ref{f25}(a) is defined by stars with $ \overline{(H-K)}_{res}$ mean values
between 0.20 and 0.24. Fig.~\ref{f25}(b) shows resulting extinctions for stars 
within 500 pc. The variation of $n_H$ with distance is not well defined, but it 
does indicate $D_{max}$ $\approx$230 pc, a value corroborated by the median 
extinction, also shown in Fig.~\ref{f25}(b), that stays constant immediately 
behind the cloud. The constancy of the median sets in at about the same 
distance $\approx$230 pc. Stars used for the distance estimate
are inscribed in diamond symbols. The estimated distance for the stars inside
the $\overline{H-K}_{res}$ = 0.20 mag contour becomes $D_{\rho ~Oph}$ = 133$\pm$6 
pc. The distance to the core region is shown in Fig.~\ref{f25}(c) and here we did not
apply a $\overline{H-K}_{res}$ criterion -- not needed anyway because any 
reseau does have a high mean $(H-K)$ value. The distance estimate
does not change $D_{LDN ~1688}$ = 134$\pm$3 pc. Finally Fig.~\ref{f25}(d) shows the
extinction jump in the southern extension, often called the arc, and both the 
$\overline{H-K}_{res}$ $>$ 0.20 and the $\sigma_{JHK} <$ 0.040 mag criterion are
applied. The solid curve is the median for the complete cloud complex and the
distance of LDN 1672 and LDN 1675 is compatible with $\sim$133 pc. We propose
accordingly that the distance to the $\rho$ Oph star forming complex is 133$\pm$6
pc not accounting for the depth of the complex.

With the advent of VLBA astrometry of low/median mass YSOs to a precision of a mere
few percent the derivation of distances to nearby star forming clouds seems 
to have entered a new era. Loinard et al. (\cite{LTMR08a}) measured parallaxes
for the two such systems, S1 and DoAr21, in LDN 1688 and found a resulting
distance they refer to as the cloud distance: 120$^{\rm +4.5}_{\rm -4.2}$ pc.
A similar distance, 119$\pm$6 pc, was suggested by Lombardi, Lada and Alves
(\cite{LLA08}) from a maximum likelyhood study of a preselected sample.
In a study of the distribution and motion of the gas in the $\rho$ 
Ophiuchi cloud from high resolution spectroscopy of Hipparcos stars
Snow, Destree and Welty (\cite{SDW08}) find a most likely distance to the dense
molecular cloud 122$\pm$8 pc and that the more diffuse component is distributed
between $\sim$110 and $\sim$150 pc.  
Knude and H{\o}g (\cite{knude98}) proposed $\sim$120 pc as the 
distance to the Ophiuchus region and suggested 150 pc as an upper limit to
the complex of clouds.

\subsection{The LDN 204 and LDN 1228 filaments}

These two filaments host four isolated 
cloud cores, Chapman and Mundy (\cite{CM2009}). Examples of cores with no YSOs
(LDN 204) and with 7 YSO candidates (LDN 1228). The two filaments are rather
nearby 125$\pm$25 and 200$\pm$50 pc as quoted by Chapman and Mundy and may 
thus be within reach of the $JHK$-photometry. Since there are three different
YSO classes in LDN 1228, Class II and earlier, a more precise distance
estimate could be useful for calibrations of PMS models.

\subsubsection{The LDN 204 filament}
The LDN 204 filament is an interesting feature because it is nearby and is 
silhoutted against the extended HII region powered by $\zeta$ $Oph$ at a
distance of only
140 pc and $\approx$ 3$^{\circ}$ away from LDN 204. The filament displays a most
regular polarization pattern and is thus a good candidate cloud for studying
the influence of the magnetic field on possible star formation. Part of the 
filament is included in an extention of the c2d study of
molecular cloud cores as a specimen of the cores presently not actively forming
stars, Chapman and Mundy (\cite{CM2009}). We might have included this cloud
under the $\rho$
$Oph$ heading since it could be part of the Ophiuchus complex of clouds as indicated
by the extinction map in Lombardi et al. (\cite{LLA08}) and it bears a certain
similarity to the appearance of Lupus I, Fig.~\ref{f30}. 

\begin{figure}
\epsfxsize=8.0cm
\epsfysize=16.0cm
\epsfbox{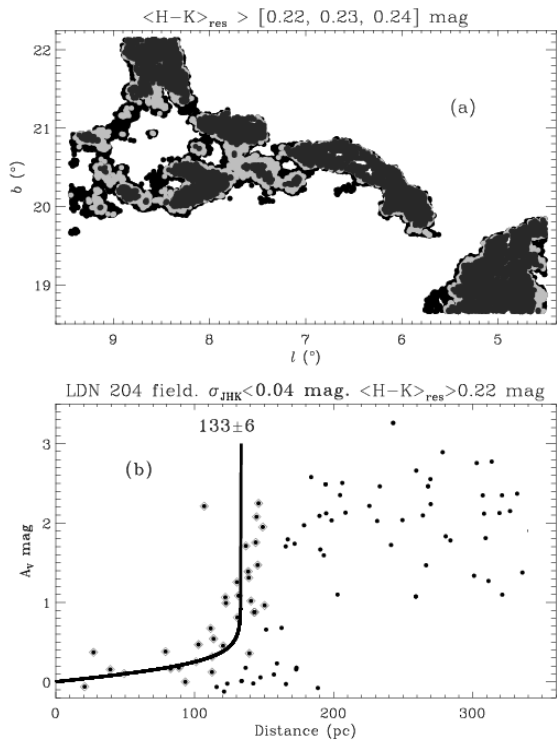} 
\caption[]{LDN 204, which does not host star formation. (a) 
Stars with $\sigma_{JHK}\leq$0.080 mag and $\overline{H-K}_{res}$ greater than 
0.22, 0.23 and 0.24 respectively. LDN 204 is the curved feature at the diagrams 
center. (b) Extinction - distance pairs for stars in reseaus with 
$\overline{H-K}_{res} >$ 0.22 mag and with $\sigma_{JHK} <$ 0.040 mag. The 
solid curve indicates the arctanh fit to the sample confined by 
$D_{max}$ = 250 pc and $D_{LDN \ 204}$ = 133$\pm$5 pc} \label{f26}
\end{figure}

The cloud outline and the extinction vs. distance may be seen in Fig.~\ref{f26}. Several
other clouds than LDN 204 appear in panel (a). We have assumed them to be spatially
associated.

The resulting distance is found as 133$\pm$6 pc identical to the distance 
suggested for the central clouds in the Ophiuchus complex. So from the distance
point of view LDN 204 and its nearest string of cloud companions seem to
belong to the Ophiuchus group of clouds.

\subsubsection{The LDN 1228 filament}

\begin{figure}
\epsfxsize=8.0cm
\epsfysize=16.0cm
\epsfbox{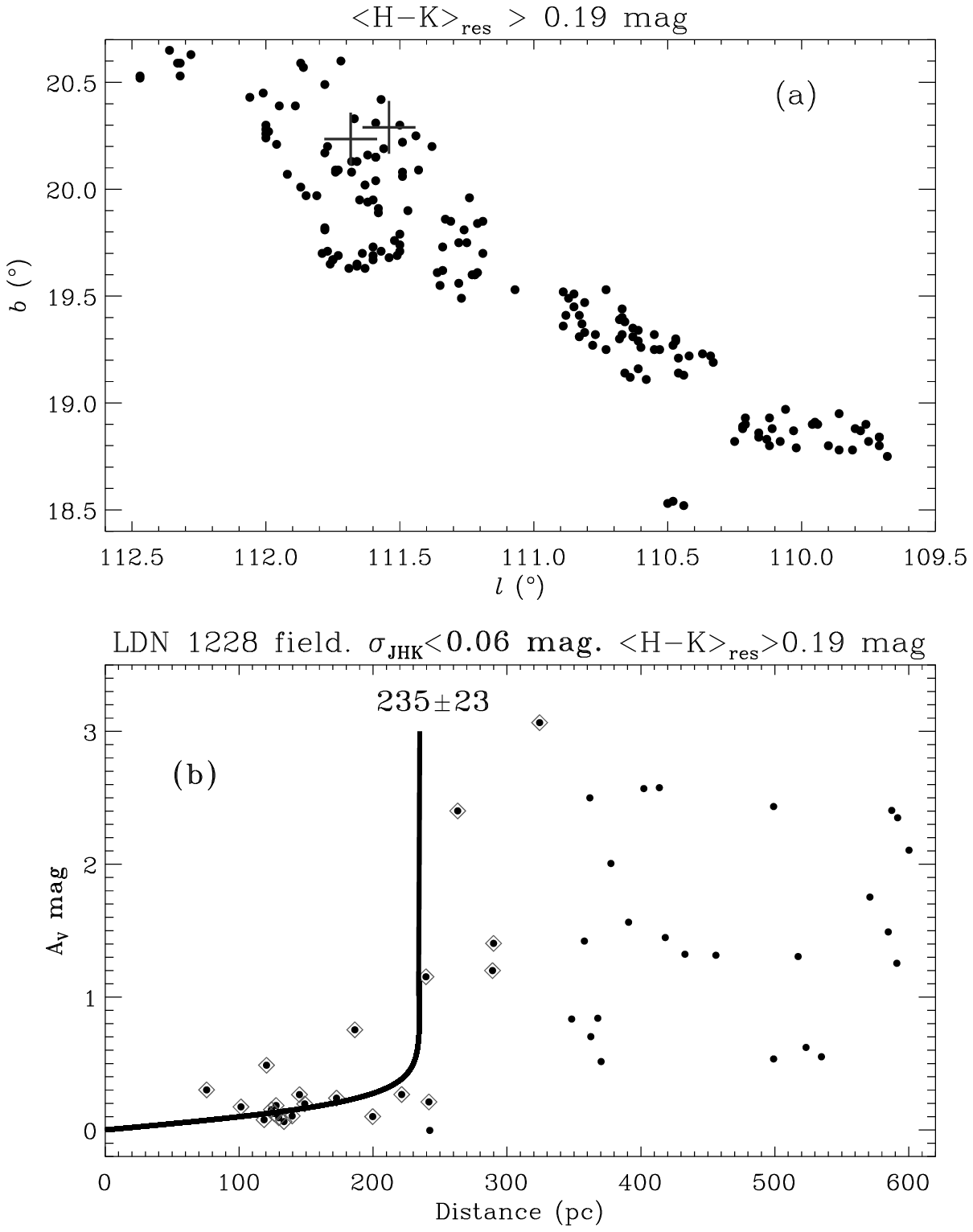} 
\caption[]{LDN 1228. (a) Collection of stars with $\sigma_{JHK}\leq$0.060 mag and 
$\overline{H-K}_{res}$ greater than 0.19. The lower left plus sign indicates the
location of HH 200 IRS and the upper right is HH 199 both of which happens
to be located at the nominal center of LDN 1228.
(b) Extinction - distance pairs for stars in reseaus with $\overline{H-K}_{res}$ 
$>$ 0.19 mag and with $\sigma_{JHK} <$ 0.060 mag. The solid curve indicates the 
arctanh fit to the sample confined by $D_{max}$ = 400 pc. $D_{LDN \ 1228}$ = 
235$\pm$23 pc} \label{f27}
\end{figure}

Chapman and Mundy (\cite{CM2009}) cite a distance 200$\pm$50 pc for this 
filament. Conelly, Reipurth and Tokunaga (\cite{CRT2008}) prefer a distance
175 pc from the compilation of LDN distances by Hilton and Lahulla
(\cite{HL1995}) formed as an average of two literature values 150, 200 pc. The
filament is known to contain HH objects within its confinement. We have taken
the nominal position ( l, b) = (111.66, +20.22) and extracted the 2mass stars
within a 4$\times$4 $\Box ^{\circ}$ area for further study. Figure~\ref{f27} (a) displays stars
with  $\sigma_{JHK} <$ 0.060 mag and $\overline{H-K}_{res}$ $>$ 0.19. The HH 199
and HH 200 positions are also shown. Note that the $\sigma_{JHK}$ criterion has 
been relaxed somewhat to have enough stars for the distance estimate. 

After applying the arctanh fit on the stars with $D_{max}$ = 400 pc
the LDN 1228 distance is estimated to 235$\pm$23 pc. The precion is inferior
to the one for the LDN 204 distance but is none the less on the $\lesssim$10$\%$
level. 

Chapman and Mundy (\cite{CM2009}) present model parameters for their YSO
candidates. If a change of distance from 200 to 235 pc applies luminosities go
up by $\approx$40$\%$. NOTE that Chapman and Mundy (\cite{CM2009}) also
suggest a
variation of the MIR extinction law; most pronounced in the possible outflow
regions but we have used our standard law despite this fact. This may be
justified by the relative shallowness of the 2mass data not probing all the way
to the PMS stars. 

\subsection{LDN 1622 and 1634 near Orion}

The Orion giant complex requires a study by itself and is not included
in the present work. We just report our results for directions towards the two 
isolated cometary clouds LDN 1622, (l, b) = (204.7$^{\circ}$, -11.8$^{\circ}$), and LDN 1634, 
(l, b) = (207.6$^{\circ}$, -23.0$^{\circ}$) both actively forming stars and possibly associated
to the Orion complex. 

\subsubsection{LDN 1622, 1621,1617, and 1624}

We have previously reported
a distance estimate based on calibrated Tycho-2 photometry and Michigan
classification, Knude et al. (\cite{KFHM2002}). In this region there is an
indication that the first dust is met somewhere between 160 and 200 pc. The use 
of the combination of Hipparcos and Michigan
classification, Fig.~6 -- 7 of Knude et al. (\cite{KFHM2002}), presents a complex
picture of the distribution of extinction with distance: we see extinction
discontinuities at approximately 160, 250 and 400 pc depending on the angular
separation from the center of LDN 1622. Due to the spatial incompleteness of the 
parallax catalog these distances, apart from the smallest one, may be due to
selection effect. The latter, however, comply with the canonically accepted Orion
complex distance.  

\begin{figure}
\epsfxsize=8.0cm
\epsfysize=16.0cm
\epsfbox{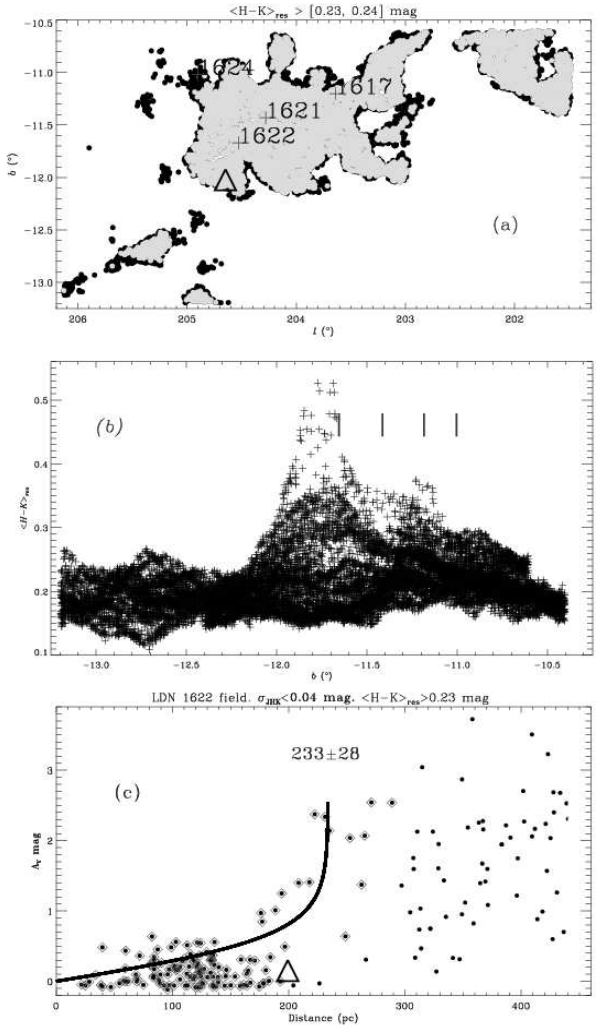} 
\caption[]{The LDN 1622 area. (a) Collection of stars with $\sigma_{JHK}\leq$0.040 
mag and $\overline{H-K}_{res}$ greater than 0.23. We have indicated the nominal 
centers of the Lynds dark clouds inside the $\overline{H-K}_{res}$ = 0.23 mag
confinement. The triangle indicates the location of HD 39572, an unreddened B9 
star located at 199$\pm^{55}_{33}$ pc as measured by Hipparcos. (b) This frame 
shows the variation of $\overline{H-K}_{res}$ with latitude including stars 
with $\sigma_{JHK}\leq$0.080 and is used for estimating the lower limit for the 
cloud confinement $\overline{H-K}_{res}$ = 0.23 mag. The vertical markers show 
the sequence LDN 1622, 1621, 1617 and 1624 (left to right). (c) Extinction - 
distance pairs for stars in reseaus with $\overline{H-K}_{res} >$ 0.23 mag and 
with $\sigma_{JHK} <$ 0.040 mag. The triangle locates HD 39572 having assumed 
that its a MS star. The solid curve indicates the arctanh fit $D_{LDN 1622}$ = 
233$\pm$28 pc to the sample confined by $D_{max}$ = 350 pc} \label{f28}
\end{figure}

We have extracted 2mass data from a $\approx$20$\Box ^{\circ}$ area with 
$\sigma_{JHK}\leq$0.080. We have chosen $\overline{H-K}_{res}$ 
$\geq$ 0.23 to represent a sight line with extinction relevant for LDN 1622. This
choice is corroborated by panel (b)  of Fig.~\ref{f28} where we have plotted 
$\overline{H-K}_{res}$ vs. latititude. Below -12$^{\circ}$ $\overline{H-K}_{res}$ is 
fairly constant and stays below $\approx$0.23 mag which accordingly is taken to
represent the maximum value valid for lines of sight outside the clouds. A relative
zero level so to say. At -12$^{\circ}$ the maximum $\overline{H-K}_{res}$ rises
dramatically. Panel (b) also displays $\overline{H-K}_{res}$ values found at the
nominal latitude of LDN 1622, 1621, 1617 and 1624 in rising order. The declining 
run of the maximum $\overline{H-K}_{res}$ 
may indicate that we are moving from the head of a cometary cloud out in its tail.
Note, however, that this is $\underline{not}$ the usual orientation of the 
cometary tail. See Fig. ~1 of Reipurth et al. (\cite{RMBW2008}) where LDN 1622's
tail is $\sim$perpendicular to the LDN 1622 LDN 1617 connection.
Panel (a) shows the distribution on the sky of reseaus with $
\overline{H-K}_{res}$ $>$ 0.23 mag. We note that LDN 1622, 1621 and 1617 are 
located along the axis of the cloud.

The cloud sample is constrained by  $\sigma_{JHK}\leq$0.040 and 
$\overline{H-K}_{res}$ $>$ 0.23. There are too few stars to use the ideal procedure
so we are obliged to use the distribution of individual values of $A_V / D_{\star}$
and we accept $D_{max}$ = 350 from this distribution. The curve fitting returns 
$D_{\it LDN \ 1622}$ = 233$\pm$28 pc for the 131 stars used in the fit. The number
of stars showing the extinction discontinuity is less than 20 as panel (c) shows.
These numbers are quite interesting considering that the 2mass extraction we
search contains more than 32000 stars. Assuming that the cloud outline in panel (a)
is due to a single structure, 233 pc may apparently also apply to 
LDN 1621, 1624 $and$ to LDN 1617. That LDN 1622 and LDN 1617 should be associated
is, however, contested by a $V_{LSR}$ difference $\gtrsim$5 km/s, Reipurth et al.
(\cite{RMBW2008}).   

Panel (c) does show a group of four stars between 170 and 200 pc showing an 
extinction $\approx$1 mag perhaps corroborating the 160 -- 200 pc estimate from 
Knude et al.  (\cite{KFHM2002}). One of the Hipparcos stars, HD 39572, with a 
measured distance of 199$\pm^{55}_{33}$ and classified as B9 is marked with a 
triangle in panel (a) and (c). Assuming that it is a main sequence star implies
 $A_V$ $\approx$0.1 mag. It is in other words not affected by the LDN~1622 
extinction. The stars position in panel (a) is inside the cloud demarcation so 
it may in fact provide a lower distance limit since it is unreddened.

\subsubsection{LDN 1634}

\begin{figure}
\epsfxsize=8.0cm
\epsfysize=16.0cm
\epsfbox{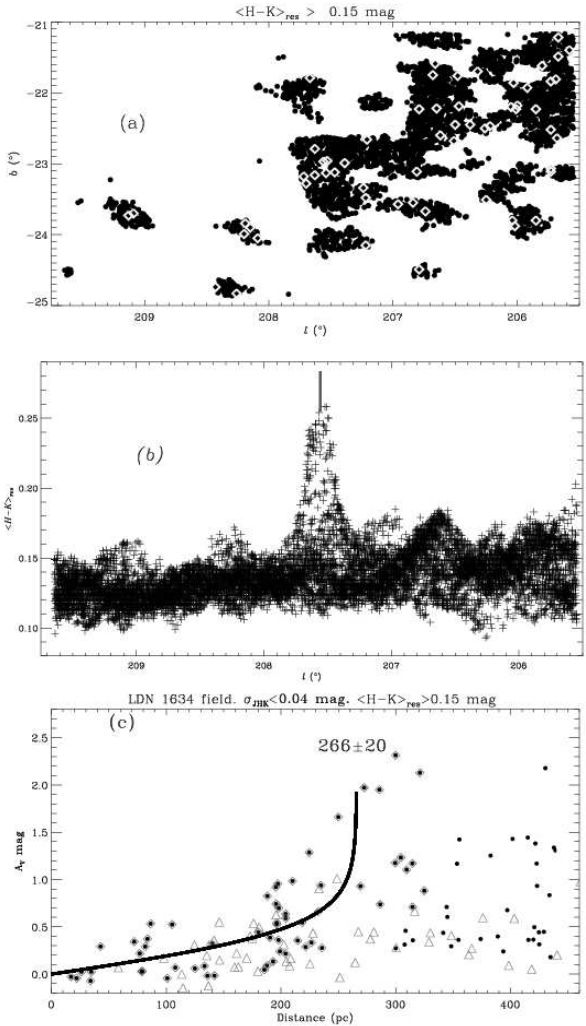} 
\caption[]{The LDN 1634 area. (a) Collection of stars with $\sigma_{JHK}\leq$0.040 
mag and $\overline{H-K}_{res}$ greater than 0.15 from a 4$\times$4 $\Box ^{\circ}$
centered on (l, b) = (207.6$^{\circ}$, -23.0$^{\circ}$). Diamonds is the sample used for the 
distance fit. (b) This frame shows the variation of $\overline{H-K}_{res}$
with longitude including stars
with $\sigma_{JHK}\leq$0.080. The vertical marker shows the longitude of LDN 1634.
(c) Extinction - distance pairs for stars in reseaus with 
$\overline{H-K}_{res} >$ 0.15 mag and with $\sigma_{JHK} <$ 0.040 mag. 
The solid curve indicates the arctanh 
fit $D_{ldn 1634}$ = 266$\pm$20 pc to the sample confined by $D_{max}$ = 425 pc.
The points inscribed in diamonds is the sample used for the distance fit. Open 
triangles are the resulting extinction -- distances variation for stars with
Hipparcos parallaxes present in the area shown in panel (a) and for which 
spectral and luminosity were available in the literature. Their extinctions are estimated independently from 2mass data.} \label{f29}
\end{figure}

LDN 1634 may resemple LDN 1622 since it is located outside the 
Barnard Loop and like LDN 1622 contains a number of young stellar objects. In
a study of these YSOs and their outflows Bally et al. (\cite{BWRM2009}) have 
estimated the clouds spatial location and its implications for its distance from 
the Sun from the influx of radiation required to keep its rim ionized. This
ionization distance
is in accord with the canonical Orion distance of 400 pc. The mass following from
a 400 pc distance implies a star formation efficiency of $\sim$3$\%$ in LDN 1634.
Fig.~\ref{f29} shows the 2mass data used for our discussion. Panel (b) 
is $\overline{H-K}_{res}$ vs. longitude and support our choice of 0.15 mag as the
lower reseau limit for the cloud lines of sight as evident for 
$l \ \gtrsim$ 208$^{\circ}$. The
distribution on the sky appears from panel (a) where we also indicate the location
of the sample used for the distance fit. Contrary to LDN 1622 LDN 1634 has a very
frayed appearence. The line of sight mean extinction, $A_V / D_{\star}$ mag/pc,
has a clear peak but is probably influenced by the presence of matter at distinctly
different distances (only three stars in fact). $D_{max}$ = 425 pc is accepted and
the fit returns $D_{LDN \ 1634}$ = 266$\pm$20 pc. The fitted curve is shown in the
(c) panel of Fig.~\ref{f29}. We have also extracted stars with Hipparcos parallaxes
from the total area in panel (a) and for those with a Michigan classification we
estimate the color excess. The variation of extinction with distance for these 
stars closer than 450 pc is shown as triangles in panel (c). Two stars at 
$\sim$250 pc in fact have an extinction $A_V$ $\approx$1 mag. So we may possibly 
maintain that some material displaying extinction exceeding what is expected from 
the diffuse ISM is found at 250 -- 266 pc. A visual inspection of Fig.~\ref{f29} 
may even suggest a distance $\approx$200 pc. This short distance estimates are 
significantly different from the detailed "ionization" distance $\sim$400 pc to 
LDN 1634 found by Bally et al. (\cite{BWRM2009}).   

\subsection{The Lupus Region}

The Lupus clouds have a complex distribution on the sky and may be
overlapping. We are therefore in need of a good confining procedure. As the
maps by Cambr{\'{e}}sy (\cite{C99}) show optical 
star counts are useful to locate clouds but the reseau average of $(H-K)$ may
be even better. 
  
Lupus ~I -- Lupus ~VI (Cambr{\'{e}}sy (\cite{C99}), form a 
complex covering a large region of the sky $\sim$10 $\times\sim$15 $\Box^{\circ}$.
The outline of the complex in integrated $^{12}CO$ intensities, $A_K$ extinction
and optical extinctions are given by Tachihara et al. (\cite{tachihara01}),
Lombardi et al. (\cite{LLA08}) and Cambr{\'{e}}sy (\cite{C99}) respectively. 

The angular extent of the
clouds alone suggests that the complex could be rather nearby. That is if
the individual clouds are physically connected. Most often these clouds are 
understood as constituting a single spatial structure. If this is the case a 
single distance applies to all constituents. A small well defined isolated 
cloud may of course have its distance given by a single number. More extended 
features may be expected to have a depth comparable to their size on the sky. 
For Lupus this would mean a depth of approximately 
2$\times$140$\times tan(\sqrt{10 \times 15} /2)$ = 30 pc. The 30 pc also 
indicates the demands on the accuracy of the estimated cloud distance. Similar 
differences may accordingly be expected between individual cloud distances.
In a detailed study of the kinematics of PMS stars in the Lupus Association 
Makarov (\cite{M07}) demonstrated that the distribution of star formation 
during the past $\approx$25 $Myr$ has had a depth of more than $\approx$30 pc. 
Roughly identical to the linear projection on the sky.  The depth of the Lupus 
complex has also resulted from a maximum likelyhood analysis of photometric 
and astrometric data for the Ophiuchus and Lupus regions, Lombardi et al. 
(\cite{LLA08}). Suggesting a thickness of Lupus of $51_{-35}^{+61}$ pc. The 
thickness likelyhood of the Lupus complex indicates that the depth may 
extend to somewhat beyond 200 pc.

With a proper distribution of stars in the $A_V ~vs. ~distance$ plane or rather
in the $H-K ~vs. ~J-H$ diagram we may obtain accuracies on the cloud distances 
from the curve fitting on the $\pm$10 pc level and may accordingly distinguish a 
cloud at $\approx$150 pc from one at $\gtrsim$200 pc. 

Apart from Lupus V the Lupus clouds have an elongated, filamentary appearence and 
are separated by regions with low or almost no extinction. Lupus I and II seem 
to be isolated from each other and from the 4 other clouds by low extinction space,
e.g. Cambr{\'{e}}sy (\cite{C99}). Since the latitude of the complex is in the 
range from $b\approx 4^{\circ}$ to $b\approx 18^{\circ}$ we may expect to have a high 
but varying stellar 
density and we may have enough stars to confine the distance interval for the curve 
fitting from the variation of the distance averaged density $n_H$ and its derivative
$\delta n_H / \delta D_{\star}$. Generally we confine the discussion to stars with
$\sigma_{JHK} <$0.040 mag. The size of the outlining $\overline{H-K}_{res}$ values
vary from cloud to cloud partly caused by the latitude range but also by the 
extinctions in the various clouds. We identify the lower limit
of $\overline{H-K}_{res}$ pertaining to the cloud sight lines from diagrams of
$\overline{H-K}_{res}$ vs. one of the celestial coordinates, see e.g. 
Fig.~\ref{f28} or ~\ref{f29}. Note that the extinctions we discuss are below 
$A_V\approx$4.5 mag. Due to the limitations of our procedure we are not able 
to measure such large extinctions as the one given for the outer contour, 
$A_V \approx$ 8 mag, in the discussion of Lupus III by Teixeira et al. 
(\cite{TLA05}).  

\subsubsection{Lupus I}

\begin{figure}
\epsfxsize=8.0cm
\epsfysize=8.0cm
\epsfbox{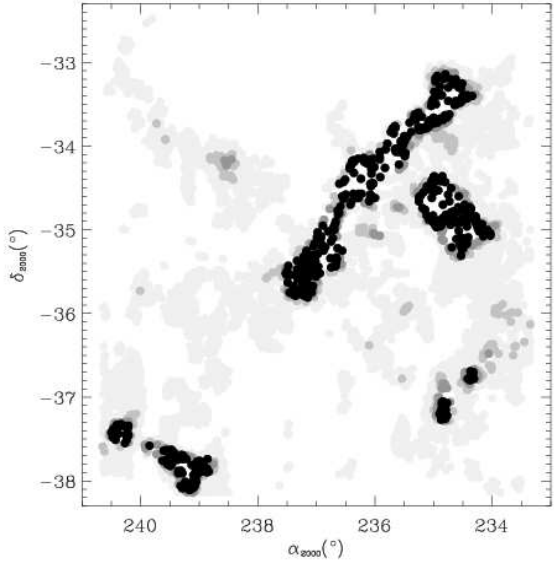} 
\caption[]{Lupus I region with an almost perfect match to the optical 
extinction map by Cambr{\'{e}}sy. $\overline{H-K}_{res}>$0.15, 0.18, 0.19 and 
0.20 mag given. The triangular feature at the lower left is Lupus II. $D_{LUP \ I}$ = 144$\pm$11 pc} \label{f30} 
\end{figure}

Fig.~\ref{f30} shows how the perimeter of Lupus I, as defined by the average 
$(H-K)$ color, changes its appearence when the lower limit is varied from 0.15 
to 0.18 whereas the appearence only changes marginally when the limit is 
raised to 0.19 or 0.20. A comparison of Fig.~\ref{f30} to the optical or 
infrared extinction maps shows a good agreement, even for minute details. 
As several other dark clouds Lupus I has low extinction arcs protruding from 
its main body.

\begin{figure}
\epsfxsize=8.0cm
\epsfysize=8.0cm
\epsfbox{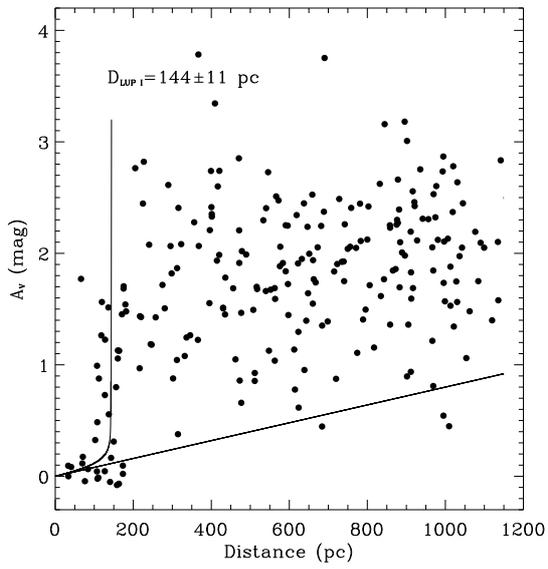} 
\caption[]{Lupus I region. Stars with $\sigma_{JHK}<0.040$ mag and inside the
$\overline{H-K}_{res}$ = 0.15 perimeter} \label{f32} 
\end{figure}

One could imagine that the $distance ~vs. ~A_V$ diagrams would depend on the 
photometric error $\sigma_{JHK}$. But applying samples with 
$\sigma_{JHK}$=0.08 and  $\sigma_{JHK}$=0.04 respectively
demonstrates that this may not necessarily be the case for Lupus I.
An eye fit of the cloud distance would indicate 100 -- 150 pc in both cases. 
The extinction rise is clearly defined by the sample 
in reseaus with $\overline{H-K}_{res}>$0.20 and $\sigma_{JHK}<0.040$. 
Confining the sample by these limitations and with $D_{max}$=250 pc in the 
$arctanh^p(D_{\star}/D_{cloud})$ fit we obtain $D_{LUP~I}$=144$\pm$11 pc as 
plotted in Fig.~\ref{f32}. In their March 10, 2008 c2d synthesis of Lupus 
Mer{\'{i}}n et al. (\cite{Merin2008}) quote 150$\pm$20, suggested by 
Comer{\'{o}}n (\cite{comeron08}) in his review of the Lupus complex, as a
reasonable Lupus I distance.
 
\subsubsection{Lupus II}
 
Lupus II is included in the lower left part of Fig.~\ref{f30} where it 
appears as an isolated feature between Lupus I and Lupus III - Lupus IV.
We have previously attempted to estimate its distance, Knude and Nielsen 
(\cite{KN01}), from V and I photometry. The $(V-I)$ distance estimate was
rather large, 360 pc, but was corroborated to some extent by Hipparcos data
for four stars, 353 pc, with a
relative precision of 30$\%$. Since the extent of the cloud is small we include
stars with $\sigma_{JHK}<0.060$ and the cloud outline is again defined by
$\overline{H-K}_{res}$ $>$ 0.20 mag. The distance fitted becomes 191$\pm$13 pc. 
Significantly larger than $D_{Lupus \ I}$ = 144$\pm$11 pc but smaller than the
$(V-I)$ estimate.

\subsubsection{Lupus III}
 
\begin{figure}
\epsfxsize=8.0cm
\epsfysize=8.0cm
\epsfbox{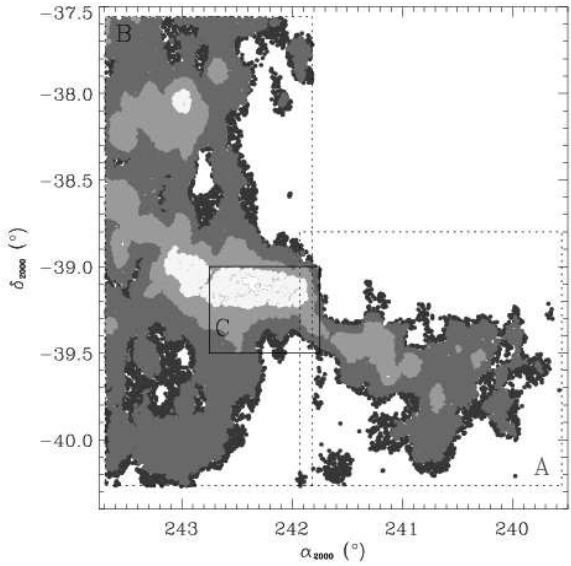} 
\caption[]{Distribution of the $\overline{H-K}_{res} ~>$ ~0.18, 0.20, 0.25, 
0.30 and 0.35 reseaus. The A, B, and C boxes for which stellar distances and 
extinctions are estimated are shown. The C-box corresponds to the region 
studied by Teixeira et al. (\cite{TLA05})} \label{f33}
\end{figure}

The projection of Lupus III shows this cloud to be one of the minor components 
of the Lupus complex and on Cambr{\'{e}}sy's extinction map Lupus III appears 
as an appendix to the apparently coherent feature consisting of 
Lupus IV, V and VI. The densest cores of Lupus III, forming the bridge head of
the filament protruding from the Lupus V and VI combination, was discussed by
Teixeira et al. (\cite{TLA05}) in a study of the physical parameters of the 
clumps with star forming activity and those without. We divide the Lupus III region
in the three subareas A, B and C indicated in Fig.~\ref{f33}. The subarea C 
contains the concentration of newly formed stars. The distance to this region is 
particularly important for estimating parameters used to study the star forming 
process.

\begin{figure}
\epsfxsize=8.0cm
\epsfysize=8.0cm
\epsfbox{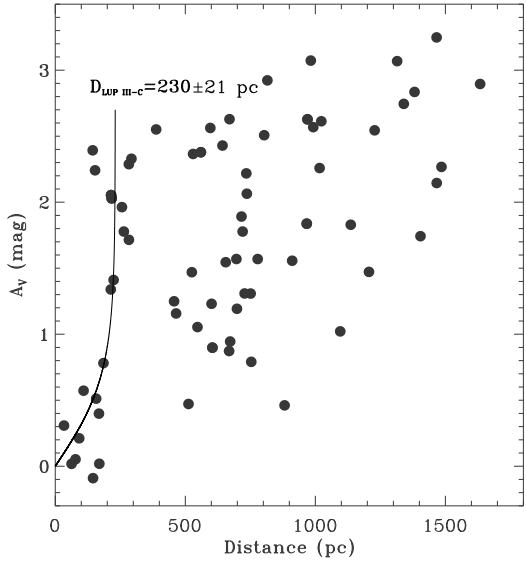} 
\caption[]{Lupus III-C region: $\alpha_{2000} \ \in [241.75, 242.75]$ and 
$\delta_{2000} \ \in [-39.5, -39.0]$. Sample limited by 
$\overline{H-K}_{res} >$ 0.18 mag and only including stars with 
$\sigma_{JHK}<0.040$ mag. $D_{LUP \ III-C}$ = 230$\pm$21 pc} \label{f34}
\end{figure}

It turns out that Lupus III as outlined by optical extinction by Cambr{\'{e}}sy
does not have a unique
distiance. The extinction contours appear to be a projection of two superposed 
clouds. The estimate of the A area is $D_{LUP ~III_A}$ ~= ~205$\pm 5$ pc compared to
$D_{LUP ~III_B}$ ~= ~155$\pm 3$ pc. The small C area covering the dense cores of 
Lupus III contains fewer stars partly because the area is small and partly due to
the larger extinction of the dense core but as Fig.~\ref{f34} shows we probably do have
enough stars for the distance estimate. The curve fitting returns the estimate
$D_{LUP ~III_C}$ ~= ~230$\pm 21$ pc in concord with $D_{LUP ~III_A}$ but 
significantly different from $D_{LUP ~III_B}$. The $D_{LUP ~III_C}$ distance is 
similar to the estimate for Lupus II of 191$\pm$13 pc. $LUP ~III_A$ and $LUP ~III_C$
are probably parts of the same physical structure located $\approx$50 pc behind the
more extended $LUP ~III_B$. The distance difference is significant on the 3 -- 5
sigma level and the distance estimates of the A+C and the B features have a
releative precision $\lesssim$5$\%$. Mer{\'{i}}n et
al. (\cite{Merin2008}) quote 200$\pm$20 again as suggested by Comer{\'{o}}n
(\cite{comeron08}) as a reasonable Lupus III distance consistent with our Lupus 
III$_{\rm A}$ and III$_{\rm C}$ estimates.

\subsubsection{Lupus IV}
Fig.~\ref{f35} displays five $\overline{H-K}_{res}$ contours of this cloud roughly
corresponding to $A_V \approx$ 0.5, 1, 2, 3, and 3.5. Fig.~\ref{f36} is showing 
distances and extinctions for stars within the $\overline{H-K}_{res}$ = 0.20 
contour. A comparison of Fig.~\ref{f35} to Fig.~\ref{f13} and Fig.~\ref{f14}, 
displaying the $\overline{H-K}_{res}$ and $count_{res}$ variation of $\sim$ the 
same region of the sky, demonstrates the better capability of the reseau colors 
to bring out the cloud perimeter. The Lupus IV data permit 
$\overline{n}_H$ and $\delta \overline{n}_H / \delta D$ to be estimated so the 
sharp definition of the sample used for the curve fitting applies. The resulting 
cloud distance from the fit $D_{LUP ~IV}$ = 162$\pm$5 pc. Lupus IV is also 
included in the c2d synthesis, Mer{\'{i}}n et al. (\cite{Merin2008}), where a 
Lupus IV distance 150$\pm$20 is quoted.

\subsubsection{Lupus V}
The projection of Lupus V is large, 4$^{\circ} \times$4$^{\circ}$ or more
and Lupus III appears as an appendix to this cloud. A major part of the cloud 
is shown in the upper panel of Fig.~\ref{f36a}. The middle panel of this 
figure shows a problem encountered when $D_{max}$ is established from the 
variation of the line of sight density and does not display a sharp peak
followed by a shallow drop off as expected from the template of Fig.~\ref{f20}(a)
but a only shallow profile without the peak. The full width distance from
a shallow profile would imply too large an estimate of $D_{max}$ again implying
too large a cloud distance. A possible interpretation of the density profile 
valid for the Lupus V area is that this cloud does not have a sharp spatial 
location but may possess a substantial depth smoothing the density peaks.
Instead we estimate $D_{max}$ from the derivative of the density or as a
slightly different approach from the derivative of the median extinction. This
latter derivative is also shown in the figure. The shape of the two derivatives
happens to be rather similar in fact. With $D_{max}$ from the half width of 
the derivatives the fitted distance to Lupus V becomes 162$\pm$11 pc. 
Interestingly Lupus III$_{\rm B}$, Lupus IV and Lupus V are at
identical distances. The nearest part of Lupus III is located at the Lupus V
distance with $D_{LP_{IIIB}}$ at 155$\pm$3 whereas Lupus III$_{\rm A}$ and 
Lupus~III$_{\rm C}$ are are found beyond 200 pc. Our estimated distances 
suggest that Lupus III$_{\rm B}$, Lupus IV and Lupus V are at a common 
distance of $\approx$160 pc. Estimating the angular diameter of the 
Lupus III$_{\rm B}$, IV and V combination to $\sim$5$^{\circ}$, e.g. from  
Cambr{\'{e}}sy's optical extinction map the projected size on the sky is
$\sim$14 pc comparable to the uncertainty $\pm$11 pc in the distance fit.
Apparently these clouds do not make up a sheet like feature.

\paragraph{Using the derivative of median extinction?}
Returning to the shallow distribution of the line of sight average density 
distribution we mentioned it possibly could be caused by a spread of Lupus V
along the line of sight which somehow contradicts the common distance of
Lupus III$_{\rm B}$, IV and V. An extinction is of course the integrated effect
of scattering and absorption along a sight line and must be related to
the intregral of the particle number density along this sight line. If we 
assume that the median extinction is representative of this integrated particle 
distribution its derivative will represent a particle density -- sort of an on 
the spot density contrary to the smooth average line of sight density. From the 
derivative of the median extinction shown in Fig.~\ref{f36a} we may possibly 
state that the corresponding density variation might be gaussian. We thus assume
that our extracted sample probes a "feature" with a gaussian number density
distribution. This "feature" is perhaps not to be perceived as a spatial 
structure since our extraction of 2mass data with distance estimate does not
probe the most dense parts of a cloud. If we assume it is located at 162 pc 
and the density distribution has a standard deviation like the uncertainty 
$\pm$11 pc. With these parameters the "feature" may 
mimic Lupus V. After integrating the gaussian distribution and assuming that the
density outside the "feature" equals the constant intercloud density the 
extinction, when scaled to the range noticed for the median extinction, becomes
as indicated in the bottom panel of Fig.~\ref{f36a}. With the assumed gaussian
density distribution the expected extinction follows the rise of the median
extinction within $\lesssim$10 pc. We are not quite sure how the result of this small
calculation should be interpreted because a single narrow gaussian does not quite
agree with the shallow variation of the average line of sight density.   

\subsubsection{Lupus VI}
Lupus VI is another example where the line of sight average density does not
display as sharp a peak as expected. Its shallow profile is evident from 
Fig.~\ref{f37} and again we use the derivative of either the density or of the 
median extinction. 
In Cambr{\'{e}}sy's extinction map the densest parts of Lupus VI seem to 
continue into Lupus IV and this is reflected in the similarity of the Lupus VI
distance 173$\pm$10 pc that does not differ from the 162$\pm$5 pc estimated
for Lupus IV. Fig.~\ref{f37} is a display of the Lupus VI data, the sample used 
for fitting a distance to the extinction jump and curves showing the variation 
of the
median density and its derivative. The curve overplotted the median extinction
has the ICM slope and may show the variation
of the extinction originating in the intercloud medium beyond Lupus VI.

\begin{figure}
\epsfxsize=8.0cm
\epsfysize=8.0cm
\epsfbox{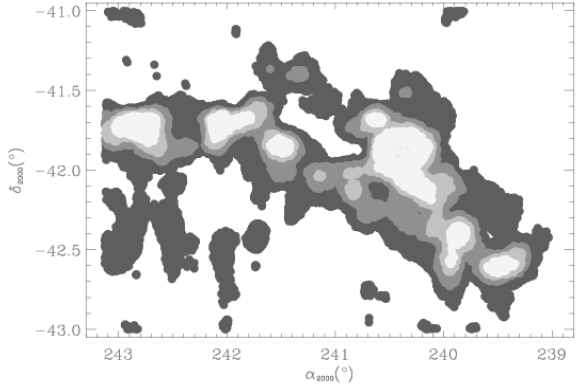} 
\caption[]{$\overline{H-K}$ contours of Lupus IV defined by 
$\overline{H-K}_{res}$ $>$ 0.16, 0.20, 0.25, 0.30, 0.35 mag inside the 
respective outlines. The boundary correspond approximately  
to $\overline{A_V} ~\gtrsim$ 0.5, 1.1, 1.9, 2.7, 3.5 mag. This diagram may be 
compared to Figs.~\ref{f13} and \ref{f14} where contours given by the reseau
counts are shown. The three eastern most clumps with 
$\overline{H-K}_{res} \ >$ 0.30 are discernible in Fig.~\ref{f14}} \label{f35}
\end{figure}

\begin{figure}
\epsfxsize=8.0cm
\epsfysize=8.0cm
\epsfbox{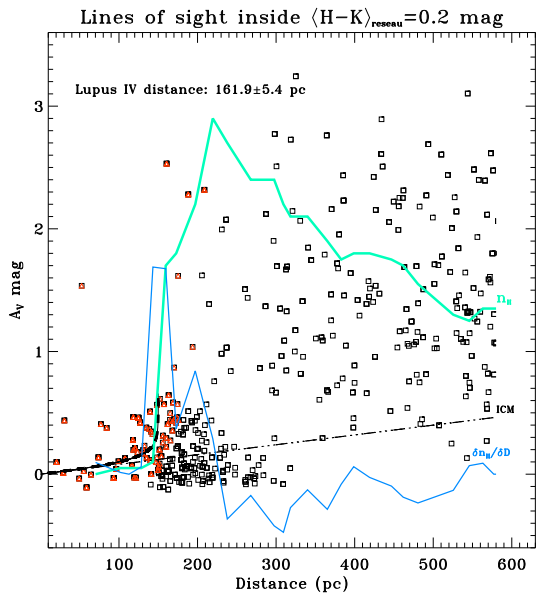} 
\caption[]{$A_V ~vs. ~D$ diagram for the Lupus IV cloud shown in the previous figure.
Arbitrarily scaled curves showing $n_H$, $\delta n_H / \delta D_{bin}$, ICM together
with the curve fitted to the extinction jump. Only stars in reseaus with 
$\overline{H-K}_{res}$ exceeding 0.20 mag. Resulting cloud distance 162$\pm$5 pc} \label{f36}
\end{figure}

\begin{figure}
\epsfxsize=8.0cm
\epsfysize=14.0cm
\epsfbox{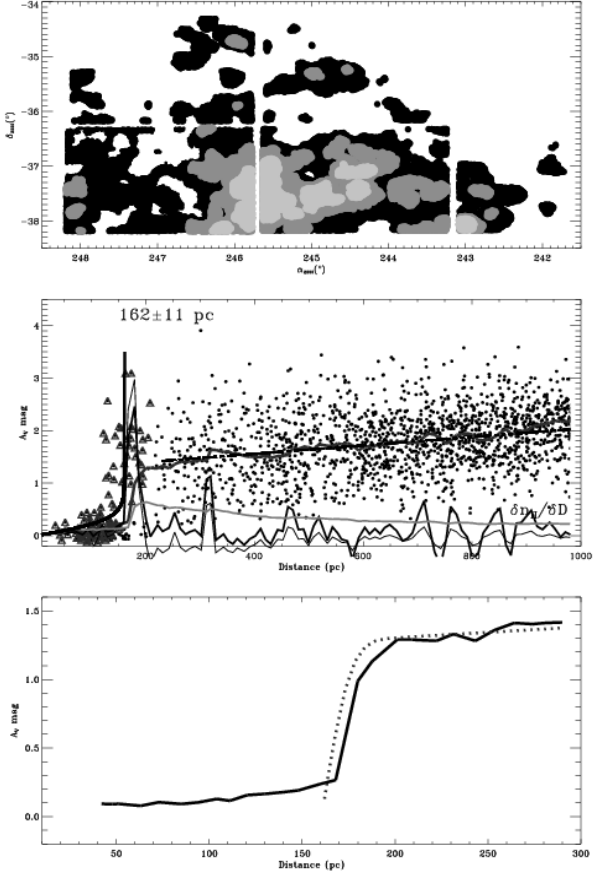} 
\caption[]{LUPUS V. $Upper panel$ area for which 2mass data with 
$\sigma_{JHK} \ <$ 0.040 have been extracted. Outer contour corresponds to 
$\overline{H-K}_{res}$ = 0.20, the next to 0.025 and the innermost to 0.30.
The $middle panel$ is the $A_V ~vs. ~D$ diagram complemented with some 
statistics. The lower thin curve is an arbitrarily scaled display of 
$\delta n_H / \delta D_{bin}$ used to estimate $D_{max}$. The lower smooth curve
is the mean line of sight density arbitrarily scaled. The median extinction
is also shown and the dashed line overplotted the median represents the ICM
variation. Only stars in reseaus with $\overline{H-K}_{res}$ exceeding 
0.20 mag equivalent to $A_V\gtrsim$1.9 mag. Resulting cloud distance 
162$\pm$11 pc is based on stars with $\sigma_{JHK} \ <$ 0.040 and located in
reseaus where $\overline{H-K}_{res} \ >$ 0.20. The $bottom panel$ shows the 
comparison of the median extinction and the extinction resulting from a 
cloud with a gaussian density distribution centered on $D_{LP V}$=162 pc 
and with $\sigma$=11 pc} \label{f36a}
\end{figure}

\begin{figure}
\epsfxsize=8.0cm
\epsfysize=14.0cm
\epsfbox{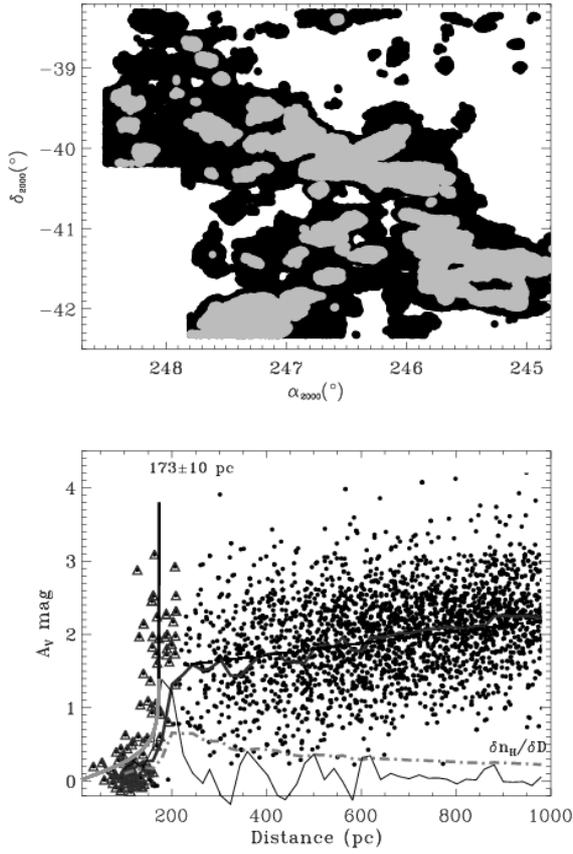} 
\caption[]{LUPUS VI. $Upper panel$ shows area for which 2mass data with 
$\sigma_{JHK} \ <$ 0.040 have been extracted. Outer contour corresponds to 
$\overline{H-K}_{res}$ = 0.25, the next to 0.030. The $lower panel$ is the 
$A_V ~vs. ~D$ diagram. The lower smooth curve is a scaled
line of sight density. The lower thin curve is an arbitrarily scaled display of 
$\delta n_H / \delta D_{bin}$ used to estimate $D_{max}$. The median extinction
is also shown and the dashed line overplotted the median represents the ICM
variation. Only stars in reseaus with $\overline{H-K}_{res}$ exceeding 
0.250 mag equivalent to $A_V\gtrsim$1.9 mag. Resulting cloud distance 
220$\pm$10 pc is based on stars with $\sigma_{JHK} \ <$ 0.040 and located in
reseaus where $\overline{H-K}_{res} \ >$ 0.25} \label{f37}
\end{figure}

\subsection{The Depth of the Lupus Complex}

\begin{figure}
\epsfxsize=8.0cm
\epsfysize=8.0cm
\epsfbox{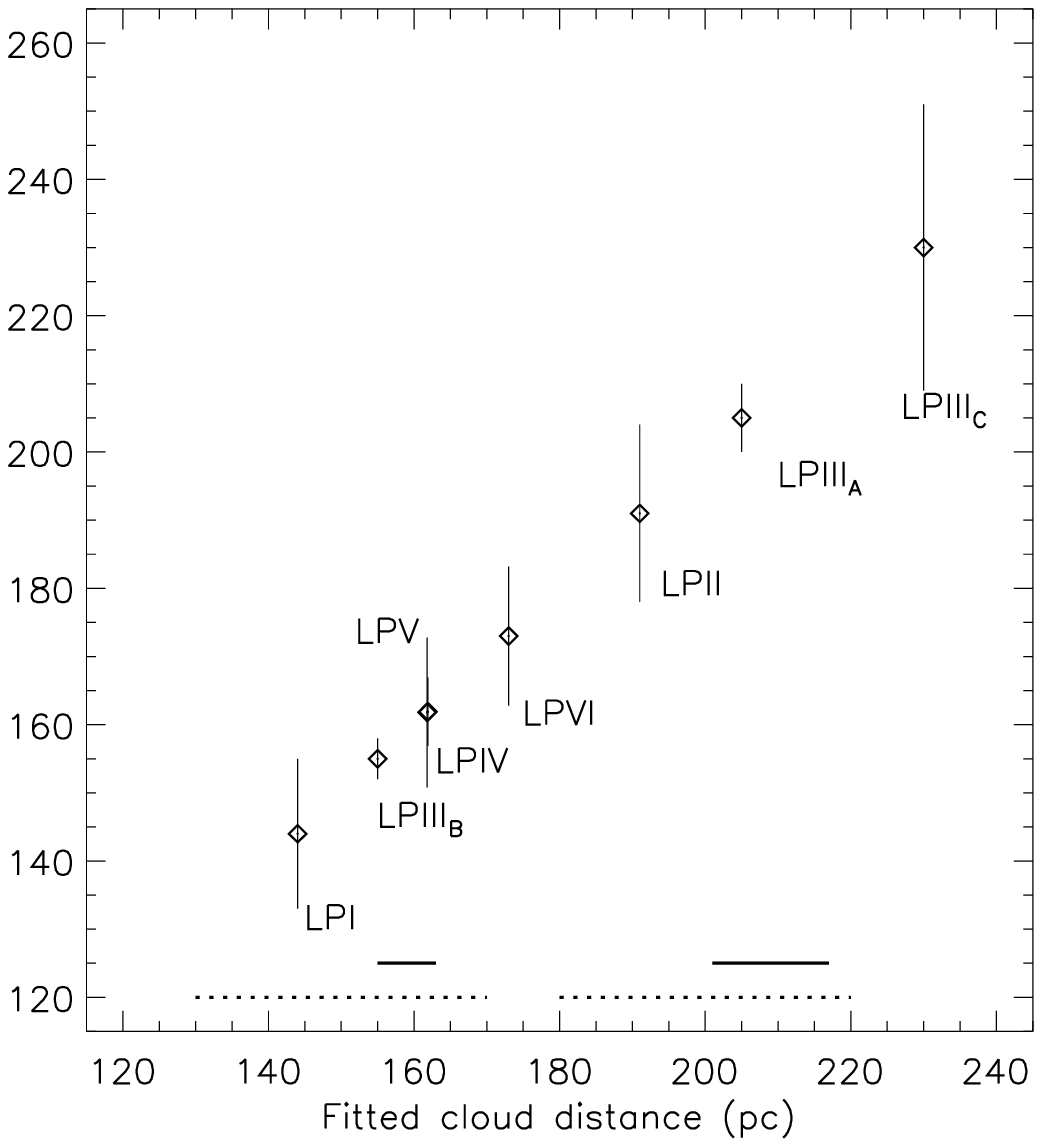} 
\caption[]{Resulting distances for the Lupus I-VI clouds with errors indicated 
for the individual cloud components. Lupus IV and V have identical distance
estimates and Lupus III is made up of components at two significantly different
distances. The dotted lines indicate the reasonable distance ranges for Lupus 
I,~IV and Lupus III suggested by Comer{\'{o}}n (\cite{comeron08}) and the ranges,
159$\pm$4 and 209$\pm$8 pc respectively, are the mean$\pm$error of the mean 
distance that we suggest for the nearby and remote group of clouds} \label{f38}
\end{figure}

The debate on the proper distance to the complex of the 
Lupus I -- Lupus VI clouds may be caused by measurements in components 
that have different distances and in particular the more shallow photoelectric 
measurements (e.g. Hipparcos and $uvby\beta$) may possibly not pertain to the 
molecular component but to the shells and sheets located in the solar vicinity. 

We have collected our distance estimates in Fig.~\ref{f38} together with the 
distance errors from the curve fitting. Apart from Lupus I, III$_{\rm B}$, V
all clouds are significantly more distant the canonical distance 140 -- 150 pc. 
Including the 8 cloud components of Fig.~\ref{f38}
the maximum cloud separation is Lupus$_{\rm depth}$ $\approx$86$\pm$24 pc. 
Excluding Lupus III$_C$ the depth
narrows to Lupus$_{\rm depth}$ $\approx$ 60$\pm$12 pc. The estimated depth is 
accordingly about three times the projected size $\approx$26 pc at 140 pc.

A simple mean of the eight distances becomes 178$\pm$3 pc where the 3
pc is the error of the mean.
At 173 pc the projected width becomes $\approx$32 pc still less than the
extent along the line of sight. In a recent review of the Lupus complex 
Comer{\'{o}}n (\cite{comeron08}) concludes that Lupus I and IV is at 
150$\pm$20 pc, Lupus III
at 200$\pm$20 pc. From our results in Fig.~\ref{f38} we notice that
$D_{LUP \ I}$=144$\pm$11 pc is compatible with a distance of 150 pc and that
$D_{LUP \ IV}$=162$\pm$5 pc seems to be marginally larger than 150. But for Lupus
III only the A and C components, see Fig.~\ref{f34}, with distances 205$\pm$5 pc and
230$\pm$21 pc are at $\sim$200 pc whereas the B component is at $\sim$150 pc. 

Taking Lupus I, III$_{\rm B}$, IV, V, VI as a common feature and Lupus II,
III$_{\rm A}$,III$_{\rm C}$ as a separate structure the first group has a mean
distance 159$\pm$4 and the second 209$\pm$8 pc consistent with the suggestion
from Comer{\'{o}}n's (\cite{comeron08}) review. Perhaps the two groups should
not be considered as spatially connected?
 
According to Tachihara et al. (\cite{tachihara01}) Lupus III 
displays the largest velocity dispersion of the Lupus I -- VI clouds indicating
a possible distribution along the line of sight.

\subsection{The Chamaeleon Clouds}

For the Chamaeleon clouds, Luhman et al. (\cite{Luhman08}),
quotes 162 pc for the Cha I distance. In Luhman (\cite{Luhman08b}) the best
Cha I distance estimate is adopted to be in the range 160 -- 165 pc. The Cha II
estimate is given as 178$\pm$18 pc adopted from Whittet et al. (\cite{W97}) 
and is marginally larger than the Cha I distance. No
estimates are given for Cha III in the review by Luhman (\cite{Luhman08b}). In 
their study of nearby molecular clouds, Knude and H{\o}g (\cite{knude98}),
detect the first indications of an extinction jump in the Chamaeleon region 
reaching $A_V$ $\lesssim$ 1 mag at a distance $\approx$150 pc based on about 10 
stars. This distance seems consistent with the 160 -- 165 pc quoted by 
Luhman (\cite{Luhman08b}). For the discussion of the Chamaeleon distance 
estimates the data and results are given in the panels of Fig.~\ref{f39}. 

\subsection{A 3$\times$3 $\square ^{\circ}$ region comprising Chamaeleon I}
Being rather nearby and accomodating active, star formation with a model median
age $\approx$2 Myr, Cha~I is a well suited cloud to search for low mass starsa
and brown dwarfs still possessing their disks, Luhman et al. (\cite{Luhman08b}),
Luhman and Muench (\cite{LM08}). Membership of the Cha I star forming clusters
was based on three criteria of which distance is just one. Distance in the 
sense that a candidate must be placed above the main sequence when shifted 
to the distance (and extinction) of Cha I. The outer contours corresponding to
$\overline{H-K}_{res}$= 0.2 mag is similar to the contours given in
Fig.~1 of Luhman and Muench (\cite{LM08}). From the variation of
$\overline{n}_{H,los}$ with distance we estimate $D_{max}$ and the arctanh 
fit returns $D_{Cha \ I}$ = 196$\pm$13 pc. Data and fit are given in 
Fig.~\ref{f39} together with Whittet et al.'s (\cite{W97}) estimate of 160 pc. 
In the Cha I frame of Fig.~\ref{f39} the filled black circles  indicate 
Whittet, Prusti and Franco, et al.'s data (\cite{W97}) and they are seen to 
follow our data closely. The 160 pc line appears as a lower distance limit to 
the jump rather than a fit. Changing the Cha I distance from 160 to 193 pc 
will increase $log \frac{L}{L_{\odot}}$ with 0.4 and as a consequence reduce the
age estimate to make it coeval to Taurus (1 Myr), Fig.~11 of Luhman 
(\cite{Luhman08b}). If the larger distance is accepted it influences our 
understanding of the disk life times.  The two distance estimates differ only 
on the 2 sigma level.

The filled circles of the Cha I and II panels of Fig.~\ref{f39} indicate the data from 
Whittet et al. (\cite{W97}) and we notice that the largest extinctions pertaining 
to the jump falls within the distance range of the stars we have used for our
curve fitting. The less extincted stars of Whittet et al. follow the ICM curve very
well.

\subsection{A 2$\times$2 $\square ^{\circ}$ region centered on Chamaeleon II}

Chamaeleon II is a nearby star forming cloud and
Porras, J{\o}rgensen, Allen et al. (\cite{PJA07}) presented Spitzer IRAC data 
for parts of Chamaeleon II where $A_V$ $>$ 2 mag. We have drawn the 2mass data 
for a similar 2$\times$2 $\square ^{\circ}$ box region centered on
$(l, b)$ = (303$^{\circ}$,-14$^{\circ}$) and with $\sigma_{JHK}$ $\leq$ 0.080 mag.

Whittet, Prusti, Franco et al. (\cite{W97}) present the photometric distance to 
Chamaeleon II as 178$\pm$ 18 pc whereas Knude and H{\o}g (\cite{knude98})
suggest 150 pc for the greater Chamaeleon region.

\begin{figure}
\epsfxsize=8.0cm
\epsfysize=8.0cm
\epsfbox{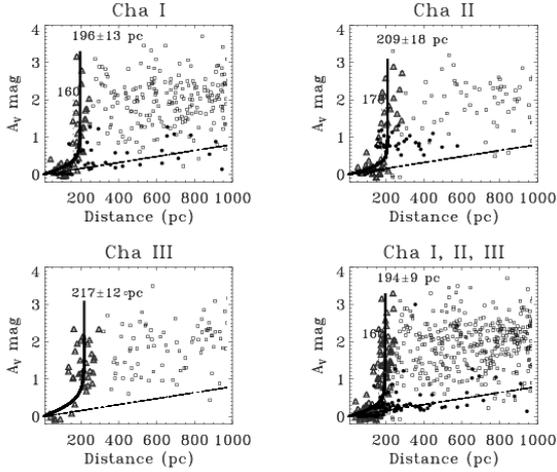} 
\caption[]{Data and distance estimates for the three Chamaeleon clouds. In the
frame to the lower right we have combined the data for the three clouds with a
resulting distance etimate $D_{Chamaeleon}$~=194$\pm$9 pc. Apart from the lower 
left panel the panels contain the photometric data (small filled circles) from
Whittet, Prusti, Franco et al.  (\cite{W97}) measured for Cha I and Cha II}
\label{f39}
\end{figure}

In the 2$\times$2 $\square ^{\circ}$ we extract stars located in reseaus
with $\overline{(H-K)}_{reseau} >$ 0.2 mag. We apply the
variation of the line of sight average density to define the stellar sample
used for fitting the arctanh function. The resulting distance is estimated to
$D_{Cha \ II}$ = 209$\pm$18 pc and is shown in Fig.~\ref{f39} together with data 
used by Whittet et al.  (\cite{W97}) for their distance 178 pc. The 178 pc almost 
appears as a lower distance limit for our cloud sample and coincides with 
$D_{fit}$-$\sigma_{fit}$ = 191 pc when we recall that in the optical gooda
individual photometric distances have a precision in the range 20$\%$ - 30$\%$. 

\subsection{The Chamaeleon III cloud}

For the sake of completeness the Chamaeleon III cloud is included
moreover because its distance has not been discussed to the same detail
as the Cha I and Cha II distances. 
Again the cloud is confined by $\overline{H-K}_{res}$= 0.20 mag but now with 
$\sigma_{JHK} <$ 0.040.
We estimate the distance to $D_{Cha \ III}$ = 217$\pm$12 pc. There is, however, a
strange lack of stars between 250 and 350 pc so the $\overline{n}_{H, \ los}$ peak
may be artificially narrow. Considering the standard deviation of the stellar 
distances the eye would probably locate the cloud at 200 pc but with the cloud
fitting sample based on $D_{max}$=350 pc from the $A_V / D_{\star}$ variation the 
fitted distances becomes slightly larger.

\vspace{0.5cm}
Since the three distances 193$\pm$13, 209$\pm$18 and 217$\pm$12 pc respectively,
are identical within the errors we combine the data with $\sigma_{JHK} <$ 0.040 
and $\overline{H-K}_{res} \ > $ 0.20 for all three clouds. The common distance
becomes 194$\pm$9 pc which is shown in the lower right panel of Fig.~\ref{f39}.
In this panel we also notice how well the minimum extinction beyond the cloud 
distance follows the diffuse intercloud extinction, $A_V$ = 0.008 mag/100 pc 
from Knude (\cite{K79}), and this includes the optical data from Whittet et al. 
(\cite{W97}) as well.

\subsection{DC300.2-16.9}

This cloud, or infrared cirrus, is located between Cha I and Cha III and Whittet 
et al.  {(\cite{W97}) assumes it is at the same distance as the Chamaeleon 
complex of clouds, $\sim$170 pc. A more recent multi-wavelength study of this 
cloud, Nehm{\'{e}} et al. (\cite{NGB08}), argues that the cloud is associated 
to the T Tauri star T Cha and that its distance accordingly is as small as the 
stellar distance of a mere 70 pc. The area inside the contour
$\overline{H-K}_{res} >$ 0.2 mag is less than one square degree and the
sample is too small to allow a good distance determination. The tail of this
cometary cloud, Nehm{\'{e}} et al. (\cite{NGB08}), extends several degrees 
towards the south and has a most patchy $\overline{H-K}_{res}$ 
distribution with 3 -- 4 apparently denser concentrations. If we relax the 
density requirement to $\overline{H-K}_{res} >$ 0.16 mag which includes 
the concentrations in the tail as well. Five stars with $A_V > 1$ mag and 
distances between 90 and 140 pc define an uprise of extinction closer than 
150 pc. Their average distance is 118$\pm$24 pc and the average extinction
amounts to $A_V$ = 1.3$\pm$0.3 mag. This is by no means conclusive but may 
indicate that DC300.2-16.9 is on the nearer side of the three Chamaeleon clouds. 
From an extensive $uvby \beta$ survey of the general Chamaeleon region 
Corradi, Franco and Knude (\cite{CFK97}) found evidence for a dusty sheet between 
100 and 150 pc which may contain the infrared cirrus DC300.2-16.9.

\subsection{The Musca cloud}

The Musca cloud is located only $\approx$4 degrees closer to the galactic plane
than Chamaeleon II and may have been formed together with the Chamaeleon clouds, 
Corradi, Franco and Knude (\cite{CFK04}), making its distance interesting to 
know. From stars in reseaus with $\overline{H-K}_{res}$ exceeding 
0.20 mag and 
with $\sigma_{JHK}$ $<$ 0.040 the resulting distance is 171$\pm$18 pc slighly 
less than the three Chamaeleon clouds, but only by a one sigma difference.

\subsection{The Southern Coalsack}

Despite the Coalsack lacks star forming activity but does contain
dense globules its distance may be interesting. Estimates of the Coalsack 
distance range from 150 to 200 pc and were summarized by 
Andersson et al. (\cite{AKS2004}).
Its location close to the galactic plane assures a high stellar density for the
extraction of usable data. We extracted data with $\sigma_{JHK}$ $<$ 0.04
mag in 9 box regions 
with centers located along the outer CO contour of 2 $K\ km \ s^{-1}$ and sides
ranging from 1.5$^{\circ}$ to 3.0$^{\circ}$. Their location and size are given in 
Table \ref{t1}.
From the distance variation of the average
density a $D_{max}$ is assigned to each sub-region and distances are estimated
in the range from 140 to 220 pc for the separate fields and they are given 
Table \ref{t1} together with their standard deviations. The 
distances are estimated on the $\lesssim$10$\%$ level. When all the data are 
combined the fitting procedure returns $D_{Coalsack}$ = 174$\pm$4 pc. The
unphysical goodness of the fit is due to the large number of available data 
points. Unphysical because the distance separation between the 140 and 221 pc
valid for the closest and remotest cloud (in our extraction) seems 
significant. With an angular diameter of 15$^{\circ}$ the estimated projected size 
is about 45 pc. There has been a discussion, based on sparse data though,
whether the Coalsack consisted of two clouds, see Andersson et al. (
\cite{AKS2004}). Whether there are two or more clouds is corroboated by the
depth noticed in the few regions we studied.

\begin{center}
\begin{table}
\caption{Distance to nine individual regions in the Southern Coalsack and the
distance, 174$\pm4$ pc, from the combined data}
\begin{tabular}{|l|c|c|c|r|r|}
\hline
Region&Size&long.&lat.&$D_{cloud}$&$\sigma_{D_{cloud}}$\\
\hline
 & $\prime \times \prime$& $^{\circ}$&$^{\circ}$&pc&pc\\
\hline
I&90$\times$90&304.5&+0.5&140&7\\
II&100$\times$100&301.5&+0.5&190&15\\
II$_{\rm b}$&100$\times$100&303.0& -1.0&208&12\\
II$_{\rm c}$&100$\times$100&303.0&+1.0&203&11\\
III&100$\times$100&304.5&-1.5&213&10\\
IV&90$\times$90&301.5&-2.5&175&12\\
V&120$\times$120&299.5&-4.0&221&13\\
VI&120$\times$120&306.0&-4.0&209&15\\
VII&120$\times$120&306.0&-1.5&160&13\\
\hline
 & & & & & \\
Combined& & & &174&4\\
 & & & & & \\
\hline
\end{tabular}
\label{t1}
\end{table}
\end{center}

Fig.~\ref{f40} shows the median extinction calculated in 20 pc wide bins with a 
50$\%$ overlap with their neighours. The 174 pc fit is also shown together 
with the range of distances displayed by the individual regions. Behind the 
extinction jump is shown the variation from the diffuse intercloud medium 
shifted by 0.5 mag. The coincidence with the median extinction may support
the presence of a void beyond the Coalsack or it may not since a $\sim$magnitude 
limited sample will fail to measure distant dense clouds.  

It is, however, interesting to compare the variation of the median extinction of the
Coalsack to the one we derived for Lupus V, Fig.~\ref{f36a} where the intercloud
slope seems applicable immediately behind Lupus V. For the Coalsack on the other
hand the intercloud slope only fits the median of the combined data $\approx$300
pc beyond the assigned distance of 174 pc and it starts at a median extinction
$\approx$0.3 mag below the peak. This may possibly be taken as an effect of a
somewhat patchy density distribution in the Coalsack, at least in the data we have
extracted, and a distribution of clouds along the sight line.

\begin{figure}
\epsfxsize=8.0cm
\epsfysize=8.0cm
\epsfbox{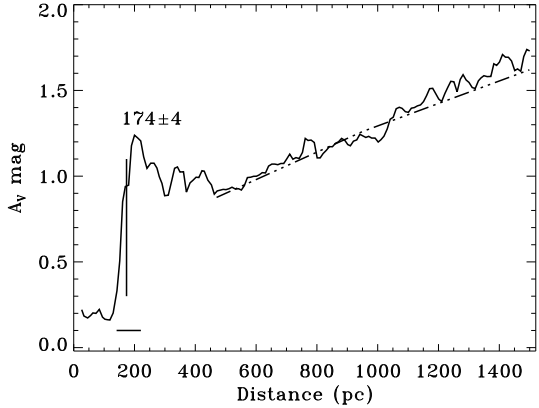} 
\caption[]{The Southern Coalsack. The solid curve shows the median extinction in
20 pc wide distance bins resulting from the combined Coalsack data. The vertical
solid line indicates the distance resulting from the curve fitting and the 
horizonthal line the range of cloud distances in the nine individual fields. The
dashed line is the expected variation caused by the diffuse intercloud medium 
shifted by 0.5 mag.}
\label{f40}
\end{figure}

\subsection{The Circinus molecular cloud complex}

The Circinus region, composed by several dark clouds, were searched for $H\alpha$
emission stars by Mikami and Ogura (\cite{MO94}) suggesting concentrations of
emission line stars at the outlines of the clouds DCld 318.8-4.4 and 
DCld 316.9-3.8 that have the largest galactic
longitudes. Compared to other molecular clouds the Circinus clouds appear much more 
frayed. An appearance ascribed to the combined effect of the outflows of previous
and ongoing star formation, Bally et al. (\cite{B99}). No dedicated efforts to 
estimate the Circinus clouds distance were found in the literature but from the
Neckel and Klare (\cite{NK80}) catalog Bally et al. (\cite{B99}) quote an extinction
increase to $A_V\sim$0.5 at $\sim$170 pc and a second jump to more than 2 mag 
between 600
and 900 pc. In Fig.~\ref{f41} we have plotted Neckel and Klare's stars from a 
5$^{\circ} \times $5$^{\circ}$ region centered on the Circinus cloud together with our
results. Distances and extinctions of Neckel and Klare's stars are mostly based on
a MK classification. We have extracted the 2mass data for five 
1$^{\circ} \times $1$^{\circ}$ regions covering the apparently densest parts of the 
complex. Confining the sample to stars located in reseaus with $
\overline{H-K}_{res}$ 
exceeding 0.35 mag and with $\sigma_{JHK}<0.040$ mag we end up with the
diagram shown in Fig.~\ref{f41}. Bally et al. (\cite{B99}) discuss the location of the
complex within 170 and 900 pc. We note that the 'wall' at 170 pc also appears in
our data beyond $\sim$200 pc but also that an extinction rise appears on the near
side of 600 - 900 pc as indicated by a $A_V$~$\approx$0.5 mag shift in the run of
the intercloud ISM, indicated by the shift of the line labelled ICM in 
Fig.~\ref{f41}.
The curve fitting suggests a 
distance 436$\pm$29 pc somewhat smaller than the 700 pc adopted by Bally et al. 
Fig.~\ref{f41} further indicates that the lower
envelope follows the general ICM slope but also that a shift of the lower envelope
may take place at $\sim$800 pc. 
436 pc is almost within the factor of 1.5 suggested as the uncertainty on the 
previously suggested distance of 700 pc, Bally
et al. (\cite{B99}). Reducing the cloud distance to 436 pc will reduce mass, linear
momentum and kinetic energy estimates by a factor $\approx$0.4 whereas dimensions 
and dynamical ages will be smaller by 
$\approx$0.6. The reduction of linear dimensions will reduce the size of all the 
outflows to $\lesssim$1 pc. More interestingly perhaps, the star formation efficiency 
given by Bally et al. will be increased by $\approx$1/0.4 implying 
$\eta_{SFE}$=3-20$\%$ counting only the four most massive stars and 
$\eta_{SFE}$=12.5-50$\%$ including the sources of all ten outflows. These 
efficiencies are rather high, the upper limits (20 -- 50$\%$) almost at the level
valid for star
forming cores, which may be right since the Circinus clouds may be remnants left
after intensive star formation. According to McKee and Ostriker (\cite{MO07}) the 
star formation efficiency is $\approx$5$\%$ averaged over the lifetime of a cloud. 

\begin{figure}
\epsfxsize=8.0cm
\epsfysize=8.0cm
\epsfbox{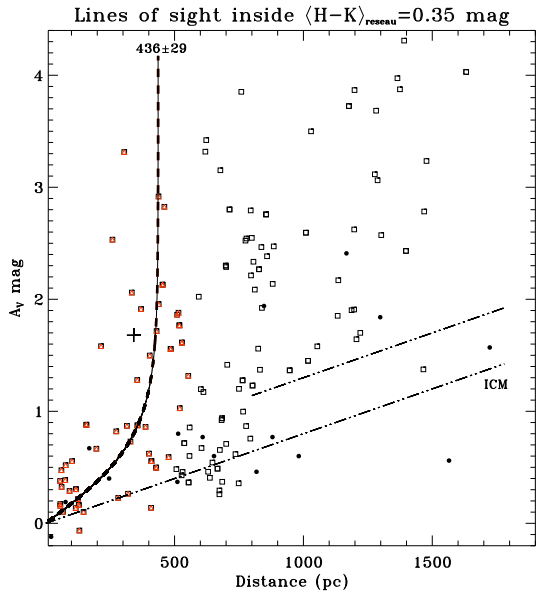} 
\caption[]{Circinus. $A_V ~vs. ~D$ diagram for a $\sim$5$\Box ^{\circ}$ region 
composed by 5$\times$1$\Box ^{\circ}$ areas covering the denser parts of the 
Circinus complex, e.g. the $H\alpha$ emission stars in DCld 318.8-4.4 and
DCld 316.9-3.8. For comparison we show the filled dots representing distance(MK)
vs. $A_V$(MK) from Neckel and
Klare (\cite{NK80}) in a larger 5$^{\circ} \times $5$^{\circ}$ region centered on the 
Circinus
cloud. Only stars with $\overline{H-K}_{res}$ exceeding 0.35 mag and with 
$\sigma_{JHK}<0.040$ are included in this diagram. For comparison Whittet's
(\cite{W07}) distance to the nearby globule DC314-05 based on a distance 
estimate of HD 130079 is indicated with a cross at 342 pc. See separate discussion
of this other cloud but with only a one sigma difference between the Circinus and 
DC314-05 distances DC314-05 could be physically associated to the 
Circinus complex?}
\label{f41}
\end{figure}

\subsubsection{DC 314.8-05.1. An isolated globule or associated to the Circinus complex?}

DC 314.8-05.1 is only removed a few degrees from the Circinus complex and may be
an example of a small isolated molecular cloud showing significant 
extinction but possibly without star formation. A particular reason for discussing
this cloud is that it , Whittet
(\cite{W07}), had it distance estimate revised from $\approx$175 pc to 342$\pm$50
pc. Due to its minor size and large extinction optical estimates of distance and
extinction may prove difficult. The 342$\pm$50 pc suggested by Whittet is based
on reflection on the dark cloud of the light from an "associated" B star and a 
larger than standard value of $R_V$=4.25 possibly justified by grain growth in
dense environments of the globule. The stellar distances we use are all based on our 
standard reddening law. The 170 pc distance estimate is again based on the 
catalog by Neckel and Klare (\cite{NK80}), see Fig.~7 of Whittet where an
extinction rise is noticed at about 200 pc, as was the case for the Circinus 
region, and a second jump at about 700 pc. From 2mass we extracted stars within 
a 1$^{\circ} \times $1$^{\circ}$ 
region centered on the globule. The extinction -- distance data are only based on
lines of sight with $\overline{H-K}_{res}$ exceeding 0.2 and 
$\sigma_{JHK}<0.04$. The interpretation of the resulting extinction -- distance 
diagram of Fig.~\ref{f42} is not simple since there are indications of two jumps 
and we may not be certain whether the apparent absence of stars
between these two jumps is real or is caused by leaving out the M0 -- M4 dwarfs.
If caused by the selection effect the distant jump should be neglected. The
first jump is at 372$\pm$52 pc and a second one at 610$\pm$25 pc. There is no sign of
the rise at 170 pc in Neckel and Klare's data, which is based on a single star 
anyway. Whittet suggests that $R_V$=4.25 for DC 314.8-05.1 and since we have been 
using the standard reddening law the use of a larger value of $R_V$ implies a shorter
distance than our suggested 372$\pm$52 pc. Comparing the DC 314.8-05.1 distance
372$\pm$52 pc from the literature to what we suggest for the Circinus complex 
436$\pm$29 pc it may not
be possible to maintain that DC 314.8-05.1 is isolated and not associated to the
nearby Circinus complex. To corroborate this possibilty we indicated Whittet's 
distance determination for DC 314.8-05.1 in the Circinus extinction -- distance
diagram, Fig.~\ref{f41}.

\begin{figure}
\epsfxsize=8.0cm
\epsfysize=8.0cm
\epsfbox{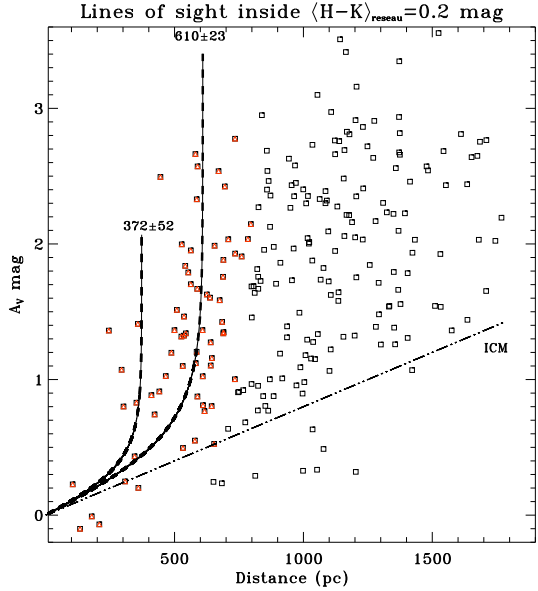} 
\caption[]{$A_V ~vs. ~D$ diagram for a 1$^{\circ} \times $1$^{\circ}$ region centered
on DC 314-05. Only stars with $\overline{H-K}_{res}$ exceeding 0.2 and 
$\sigma_{JHK}<0.040$. $D_{DC \ 314-04}$ = 610$\pm$25 pc. Whittet's 
(\cite{W07}) distance based on HD 130079, that is reflected on DC 314.8-05.1, 
amounts to 342$\pm$50 pc}
\label{f42}
\end{figure}

\subsection{IC 5146. A more distant cluster and cloud}

Extinction and molecular gas in a dark cloud near the cluster IC 5146 was
discussed in a seminal paper by Lada et al. (\cite{LLCB94}) and the cloud structure
was investigated in more detail by a deeper H, K survey suggesting extinctions 
above $A_V \sim$20 mag, Lada, Alves and Lada (\cite{LAL99}). The distances to 
the cloud and cluster, which can not be assumed to be identical {\it a priori},
are of some interest since the molecular filament studied
by Lada et al. shows a
$\sigma_{A_V}$ vs. $A_V$ variation and that the volume density falls off like 
$r^{-2}$ over scale lengths in the range 0.07 $-$ 0.4 pc assuming a distance 
of $\sim$460 pc. The $\sigma_{A_V}$ vs. 
$A_V$ variation was shown to be a consequence of the volume density variation 
$\sim r^{-2}$ and not a result of the supersonic tubulence model proposed by 
Padoan, Jones and Nordlund (\cite{PNJ97}). The young cluster IC 5146 
contains a multitude of $H_{\alpha}$ emission stars, Herbig and Dahm 
(\cite{HD02}). Despite an angular separation $\approx$1.3$^{\circ}$ on the sky, 
see Fig.~\ref{f43}, it has been assumed that filament and cluster are at the 
same distance. The filament distance was first asumed to $\approx$400 pc by 
Lada et al. (\cite{LLCB94}) mainly due to a
lack of foreground stars to the filament. In Lada et al. (\cite{LAL99}) the 
estimate was changed to 460 pc with a one sigma range from 400 to 500 pc. These 
estimates are about half the distance estimated to the cluster IC 5146. Herbig and 
Dahm (\cite{HD02}) adopt what they term a compromise distance to the cluster of 
1.2 kpc from estimates ranging from 0.9 to 1.4 kpc and quotes an uncertainty 
$\pm$180 pc coming exclusively from the uncertainty of the $M_V$ calibration of 
the three B8, B9 stars used to locate the $V_0 ~vs. ~(V-I)_0$ locus pertaining to 
the Pleiades and thought to represent the IC 5146 main sequence as well. Harvey et 
al. (\cite{H2008}) use a similar technique on B type stars projected on the 
cluster area and evaluate a new photometric distance by replacing the 
Schmidt-Kaler ZAMS by a newer luminosity calibration with data from the $\lesssim$1 Myr 
Orion Nebula Cluster which recently had a precise VLBA distance determination. 
Seven B-type stars are available, two were discarded on the grounds that they gave 
distances in the 300 -- 400 pc range. Five late B-type stars provide an average 
distance module 9.89 mag with a standard error 0.18 mag implying the estimate 
950$^{+83}_{-75}$ pc for the IC 5146 cluster.   

\begin{figure}
\epsfxsize=8.0cm
\epsfysize=8.0cm
\epsfbox{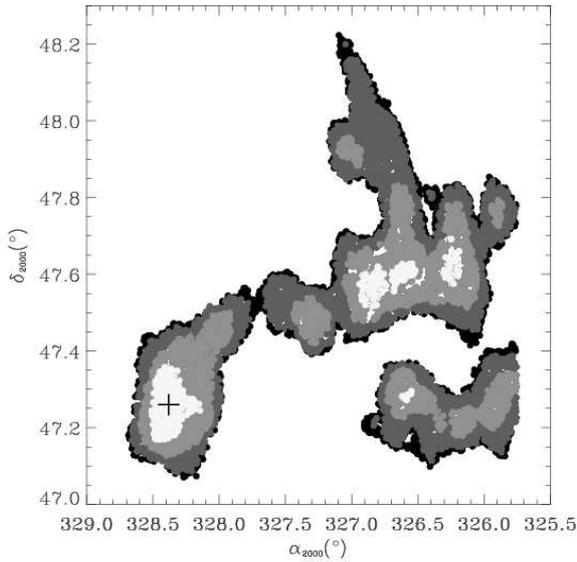} 
\caption[]{IC 5146. Projection of $\overline{H-K}_{res}$ =  
0.21, 0.26, 0.31 and 0.34 mag respectively for the two areas searched for 
stars that may be used for the distance estimate. The plus sign indicates the 
position of the IC 5146 cluster. The filament approximately centered at 
$\delta \approx$ 47.5 and $\alpha$ in the range from 326 to 327$^{\circ}$ is the 
cloud discussed by Lada et al. (\cite{LLCB94}). Note that the angular separation
between this filament and the IC 5146 cluster is about 1.3$^{\circ}$}
\label{f43}
\end{figure}

\begin{figure}
\epsfxsize=8.0cm
\epsfysize=8.0cm
\epsfbox{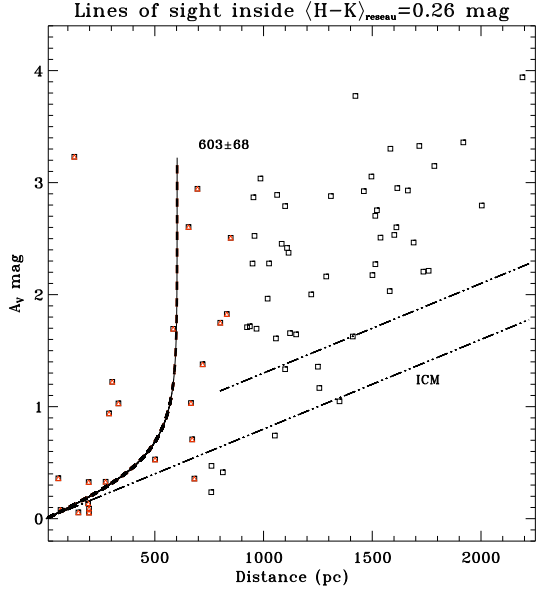} 
\caption[]{IC 5146, northern filament. $A_V ~vs. ~D$ diagram from a
1.3$^{\circ} \times $1.3$^{\circ}$ region centered on the part of the IC 5146 clouds 
studied by Lada et al. (\cite{LLCB94}) where the NICE extinction estimate was 
introduced. Only stars with $\overline{H-K}_{res}$ exceeding 0.26, with 
$\sigma_{JHK}<0.050$ and $\alpha_{2000}<327.5^{\circ}$ are included in the 
distance fit. A distance of 603$\pm$68 pc results}
\label{f44}
\end{figure}

\begin{figure}
\epsfxsize=8.0cm
\epsfysize=8.0cm
\epsfbox{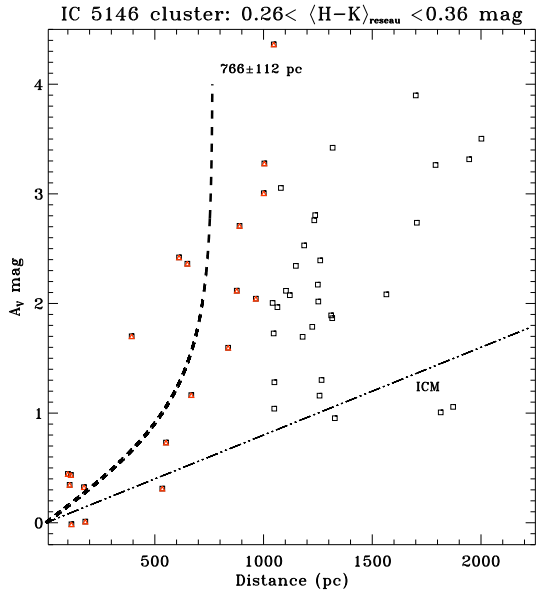} 
\caption[]{IC 5146 cluster. $A_V ~vs. ~D$ diagram from a 1.3$^{\circ} \times 
$1.3$^{\circ}$ region
containing the IC 5146 cluster. Only stars with $\overline{H-K}_{res}$ 
exceeding 0.26 but less than 0.36 and with $\sigma_{JHK}<0.050$ are included.
The upper limit is introduced to avoid the cental parts of the cluster
and $\sigma_{JHK}<0.050$ are included. Only stars with $\alpha_{2000}>327.5^{\circ}$
included. The distance of 765$\pm$112 pc 
results from fitting the $arctanh^p (D_\star /D_{cloud})$}
\label{f45}
\end{figure}

We have extracted 2mass data for the area shown in Fig.~\ref{f43} and used the 
reseau mean of $(H-K)$ to indicate the extinction contours. In order to have 
enough stars for the reseau mean values we use stars with $\sigma_{JHK}<$0.1 mag 
somewhat larger than our preferred choice of 0.04 mag. The IC 5146 cluster
position is indicated by the plus sign. From the mean color contours it is not 
obvious that the cloud filament and the cluster are parts of a coherent dust 
structure. Only a minor change of $\overline{H-K}_{res}$ from 0.20 to 0.21 
breaks the color bridge from the cloud to the cluster. For the northern 
filament we extract stars with $\sigma_{JHK}<$0.05 mag and 
$\overline{H-K}_{res}>$0.26 and $\alpha_{2000}<327.5^{\circ}$.
Distance-extinction pairs included for the curve fitting 
have a limiting distance $D_{max}$ of 1000 pc. There are too few stars
to define the distance range of the extinction rise from the variation of the mean
density vs. distance. From Fig.~\ref{f44} we notice how well the lower envelope is 
represented by the increase caused by the diffuse "intercloud" medium. The filament
distance resulting from the fit is 603$\pm$65 pc corresponding to a $\approx$10$\%$
accuracy. The suggested distance to the cloud filament is roughly $2\sigma$ above
the distance range 400$-$500 proposed by Lada et al. (\cite{LAL99}). The scale 
length will change from 0.07$-$0.4 pc to 0.09$-$0.5 pc. Mass estimates will 
increase almost by a factor of 2 ($\sim$1.7) if the increased distance estimate of 
603 pc is accepted.

The area used to study the IC 5146 cluster region has $\alpha_{2000}>327.5^{\circ}$ 
and is confined to 0.26$<\overline{H-K}_{res}<$0.36. The upper confinement is chosen
to avoid the inner parts of the cluster region where dust modifications may haven
taken place and the colors may be influenced by warm dust emission. Fig.~\ref{f45} 
shows the extinction vs. distance digramme for the "outer" parts of the cluster 
region. The fitting sample was limited by $D_{max}$ = 1200 pc, increasing $D_{max}$
to 1300 pc did not change the estimated distance 766$\pm$113 pc. The relative 
distance error is now gone up to 14$\%$ - really not bad for a feature possibly 
located at $\sim$0.75 kpc.

The distance discrepancy between the northern dark cloud and the
cluster remains but is narrowed
from 460-1200 pc to 603-766 pc. The difference of our estimates is significant
on the 2$-$3 sigma level. Taken at their face value and with a separation of 
1.3$^{\circ}$ the filamentary cloud and the cluster will be separated by 
$\approx$163 pc and may accordingly not be physically related. Conversely
Harvey et al. (\cite{H2008}) use circumstantial evidence to argue that cloud 
and cluster are at similar distances. 

\subsection{The Corona Australis Cloud}

Compared to other star forming clouds Corona Australis has an isolated location
at $\sim$18$^{\circ}$ below the galactic plane and may have another formation
history than most molecular clouds, Neuh{\"{a}}user and Forbrich (\cite{NF08}).
We have previously estimated the distance to the Corona
Australis Cloud, Knude and H{\o}g (\cite{knude98}), using Hipparcos parallaxes
and color excesses including stars within 5$^{\circ}$ from $(l, b)$ = (360.0, -20)
and noticed a marked rise in the color excess at $\sim$170 pc present in 
$\approx$15 stars with an estimated $A_V$ range from 0.1 to 1.0 mag. In their
isodensity maps of the local bubble Lallement et al. (\cite{LWVCS03}) indicate
a location of the CrA cloud at $\approx$120 pc. Three late B-type stars are
close to the projection of the denser parts of the cloud and Neuh{\"{a}}user
and Forbrich (\cite{NF08}) suggest their Hipparcos parallaxes for a weighted mean
130 pc as the Corona Australis distance.  

\begin{figure}
\epsfxsize=8.0cm
\epsfysize=16.0cm
\epsfbox{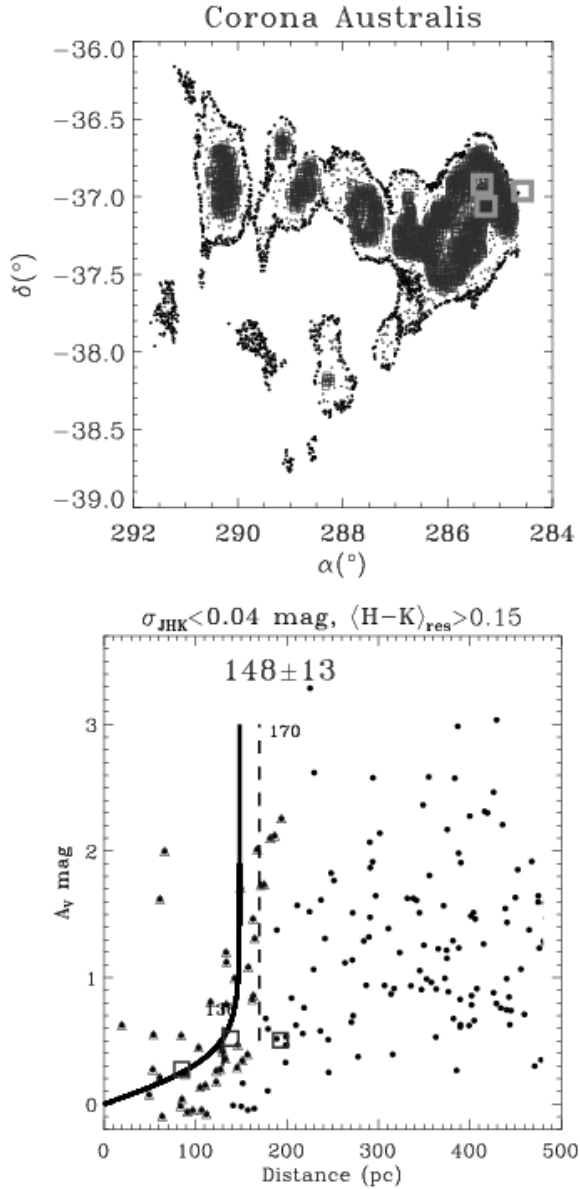} 
\caption[]{In the $upper$ $frame$ the Corona Australis cloud when confined by
$\sigma_{JHK}<0.040$ and 0.15$<\overline{H-K}_{res}<$0.16. The darkest points 
indicate the denser reseaus with $\overline{H-K}_{res}>$0.20.
The $lower$ $panel$ shows the extinction -- distance variation 
and the resulting estimate of the Corona Australis cloud distance 
$D_{cloud}$=148$\pm$13 pc. The dashed line at 170 pc is from Knude and 
H{\o}g (\cite{knude98}). The 
three squares represent the three B-type Hipparcos stars used by
Neuh{\"{a}}user and Forbrich (\cite{NF08}) for a cloud distance 130 pc} 
\label{f46}
\end{figure}

We have extracted 2mass data for this cloud with $\sigma_{JHK}<0.040$ and limit
the study to stars with $\overline{H-K}_{res}>$0.15. The extraction with 
0.15$\le \overline{H-K}_{res}<$0.16 is shown in Fig.~\ref{f46} and is in fact a rather
good representation of the optical extinction map from Cambr{\'{e}}sy 
(\cite{C99}). With $D_{max}$ = 250 pc the resulting distance $D_{CRA}$ = 
148$\pm$13 pc. Fig.~\ref{f46} also contains our previous estimate of 170 pc directly
from the Hipparcos parallaxes and the location 130 pc recently suggested.
The three stars on which the 130 pc distance is based display a low extinction and
follow the general trend based on the 2mass data but is possibly underestimating
the distance. Their relevance for the Corona Australis distance originates 
from the fact that they are likely to be CrA members.

\subsection{LDN 1450, HH 7 - 11 or NGC 1333 in the Perseus Cloud}

The dark cloud associated with the reflection nebula NGC 1333 hosts a number of pre
main sequence stars, some even of the earliest classes 0 and 0/I, according to 
several authors, e.g. Chen, Launhardt and Henning (\cite{CLH2009}), Winston, 
Megeath, Wolk et al. (\cite{WMW2009}). Ages of these PMS stars range from 1 to 10 
Myr with most objects being younger than 3 Myr. LDN 1450 appears to be associated
to a complex of dark clouds reaching all the way to IC 348 -- the Perseus Cloud.
We have extracted 2mass data from a 4$\times$4 $\Box ^{\circ}$ area centered on 
$(l, b)$ = (158.3$^{\circ}$, -20.5$^{\circ}$). We only include stars located in 
a reseau with $\overline{H-K}_{res}$ $>$ 0.20 mag and $\sigma_{JHK}$ $<$ 0.040 
mag. The distribution of lines of sight for which a distance -- extinction pair 
could be computed is shown in Fig.~\ref{f47}(a) and the pairs displayed in the 
(c) panel together with the resulting estimate of the cloud distance 
$D_{LDN \ 1450}$ = 213$\pm$12 pc. $D_{max}$ = 350 pc because the average line 
of sight density shows a rather wide distribution 
implying the large value of $D_{max}$. There exist an earlier estimate of the
LDN 1450/NGC 1333 distance from Vilnius photometry in an area comparable to the
one studied presently, $\check{C}$ernis (\cite{CERNIS1990}). The Vilnius data are 
given for comparison 
in Fig.~\ref{f47}(b) and as smaller triangles in the (c) panel. The distance proposed
from the Vilnius data is 220$\pm$20 pc and results from a weighting scheme 
including the most remote stars with $A_V$ $<$ 0.7 mag and the nearest ones with 
$A_V$ $>$ 1.5 mag. The agreement between the present estimate of 213 pc and the
Vilnius estimate of 220 pc is certainly acceptable. 

The distance to a group
of masers associated to SVS 13 in NGC 1333 has recently been obtained from multi
epoch VLBI interferometry and is reported as 235$\pm$18 pc, Hirota et al. 
(\cite{HIROTA2008}). Chen, Launhardt and Henning (\cite{CLH2009}) prefer a distance
350 pc for consistency with the literature but reduction of the distance with a
factor 223/350 might
influence the deduction of the protostar parameters and the separation of the 
components of the binary protostar in SVS 13 B subcore, as discussed in Section 
4.4 of Chen, Launhardt and Henning (\cite{CLH2009}). Knowing precise linear 
dimensions in a cloud is of course of some relevance for the discussion of 
rotational and orbital energies. 

\begin{figure}
\epsfxsize=8.5cm
\epsfysize=17.5cm
\epsfbox{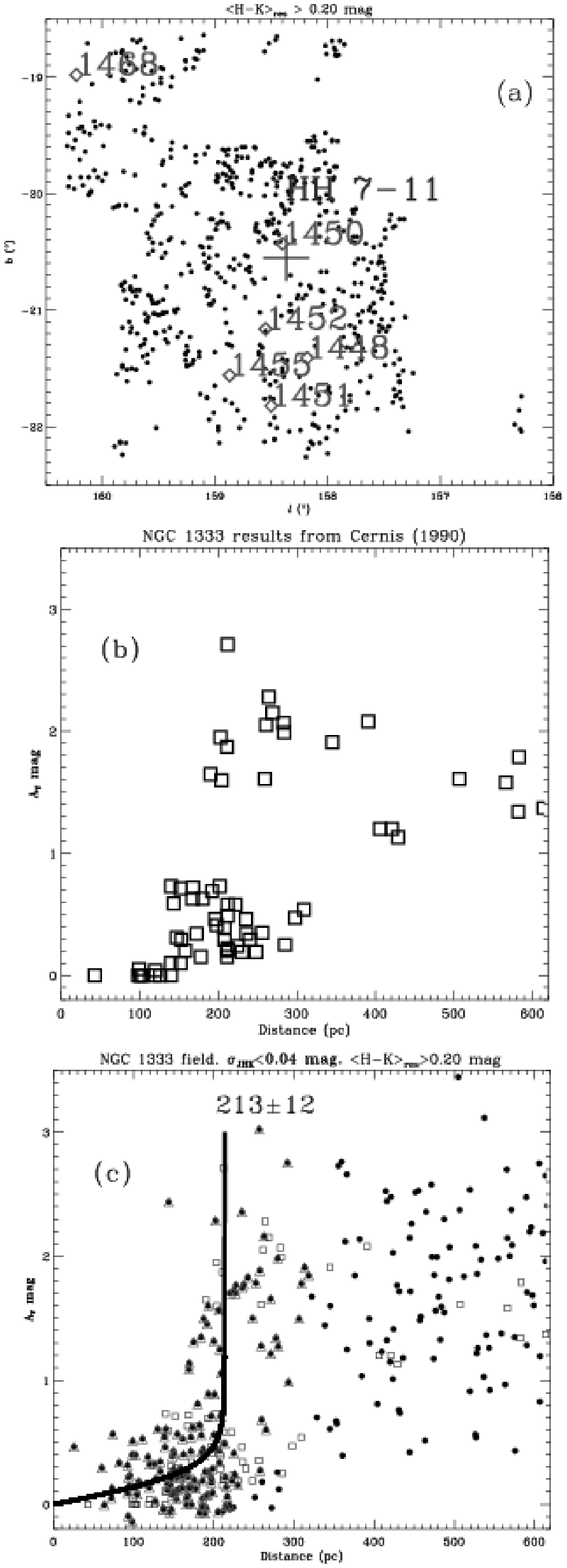} 
\caption[]{NGC 1333 in the Perseus complex. (a) Collection of stars with 
$\sigma_{JHK}\leq$0.040 mag and $\overline{H-K}_{res}$ greater than 0.20 from 
a 4$\times$4 $\Box ^{\circ}$ region centered on the NGC 1333 position $(l, b)$ = 
(158.3$^{\circ}$, -20.5$^{\circ}$). The cross indicates HH 7-11 or SVS 13 for which an 
accurate parallax has been established, 
Hirota et al. (\cite{HIROTA2008}), $D_{SVS 13}$ = 235$\pm$18 pc. 
Other numbers designate dark nebulae from the Lynds catalog. (b) Resulting 
extinction -- distance pairs from $\check{C}$ernis (\cite{CERNIS1990}) derived 
from Vilnius photometry. Dust was suggested at 160 pc and at 220$\pm$20 pc
from these data. (c) Extinction - distance pairs for stars in reseaus with 
$\overline{H-K}_{res} >$ 0.20 mag and with $\sigma_{JHK} <$ 0.040 mag. The 
solid curve indicates the arctanh fit to the sample confined by $D_{max}$ = 
350 pc. Small triangles are the Vilnius data given for comparison}
\label{f47}
\end{figure}

\subsection{The California Molecular Cloud}

The cloud containing NGC 1333 is part of the complex of clouds termed the Perseus
Cloud. It has recently been realized that the sky close to the Perseus and 
Taurus-Auriga complexes contains a major molecular, coherent cloud, Lada, Lombardi
and Alves (\cite{LLA2009}). The location 
has been known to contain a string of Lynds dark clouds. The new cloud is termed
the California Molecular Cloud and with a distance 450$\pm$23 pc suggested by Lada,
Lombardi and Alves it aspires to be
a giant molecular complex of a $\sim$10$^{\rm +5}$ M$_{\odot}$ mass and a linear 
extent of $\sim$80 pc.   

In their derivation of the distance 450 pc Lada, Lombardi, and Alves
(\cite{LLA2009}) quotes previous photometric distance estimates to dust layers at
distances 125 and 300 -- 380 pc, Ekl{\"{o}}f (\cite{E1959}) but suggest that these
layers may not be associated with the California Molecular Cloud (CMC) but rather
have their origin in Taurus-Auriga and Perseus complexes at $\approx$140 and 
$\approx$250 pc and thus falls short of the 450 pc. 

The results from Ekl{\"{o}}f (\cite{E1959}) are based on blue and red photographic
photometry and spectral classification from Schmidt plates of 1800 stars in the
Auriga region.

We estimate distances 147$\pm$10 and 213$\pm$12 pc for Taurus and Perseus
respectively, see Table~\ref{t2}. Since CMC may rival the Orion giant molecular clouds as
the most massive cloud within 0.5 kpc its distance is of interest and it is included
in the present study. Fig.~\ref{f48}(a) shows the outlines of the cloud indicated by reseaus
with $\overline{H-K}_{res}$ $>$ 0.23 and 0.28 (small $\diamond$s) respectively. The large
diamonds of panel (a) display the location of the two strings of Lynds dark clouds 
and also the location of NGC 1579 ($\bigtriangleup$). Panel (b) is the resulting    
variation of extinction with distance for the same two samples. From the 
$\overline{H-K}_{res}$ $>$ 0.23 mag, $\sigma_{JHK}$ $<$ 0.040 mag sample with 
$D_{max}$=500 pc from FWHM of the $A_V / D_{\star}$ variation. The distance of CMC
becomes 330$\pm$5 pc.  

A closer inspection of Fig.~\ref{f48}(b) reveals an apparent absence of stars between
$\approx$200 and $\approx$300 pc and with $A_V$ ranging from about 1 to about 2 so
it appears that there is a cloud layer in front of the CMC proper. Assuming that
this layer is connected to the Perseus cloud and not to the CMC layer we may
correct for its influence on the
distance estimate by removing the stars indentified in a way similar to identifying
the sample used to estimate $D_{NGC \  1333}$ = $D_{Perseus}$ = 213 pc by using
$D_{max}$ = 350 pc and remove these stars, supposed to belong to a Perseus layer of
clouds, from the $\overline{H-K}_{res}$ $>$ 0.23 mag, $\sigma_{JHK}$ $<$ 0.040 mag
$D_{max}$=500 pc sample. The CMC estimate is thus raised to 362$\pm$3 pc shown in
Fig.~\ref{f48}(b) as the thin solid curve.

CMC is an example of a cloud where we may overestimate the number of M4 -- T tracers
because we mistake O -- G6 extincted by more than 6 -- 3 mag for less reddened
late type dwarfs. We have therefore tried to exclude the M4 -- T stars from the
$\overline{H-K}_{res}$ $>$ 0.23 mag, $\sigma_{JHK}$ $<$ 0.040 mag $D_{max}$=500 pc 
sample. $D_{CMC}$ now becomes 328$\pm$4 pc as shown in Fig.~\ref{f48}(c).

We suggest accordingly that $D_{CMC}$ is between 330$\pm$5 pc and 362$\pm$3 pc or
roughly 100 pc less than estimated by Lada, Lombardi, and Alves (\cite{LLA2009}).
Interestingly this distance range is within the distance limits suggested by 
Ekl{\"{o}}f (\cite{E1959}) for the second cloud layer in his Auriga survey, 
300 -- 380 pc. The smaller distance will cause a decrease of the linear extent 
to $\approx$60 pc and of the mass to $\sim$10$^{\rm +4.73}$ M$_{\odot}$.  

\begin{figure}
\epsfxsize=8.5cm
\epsfysize=14.5cm
\epsfbox{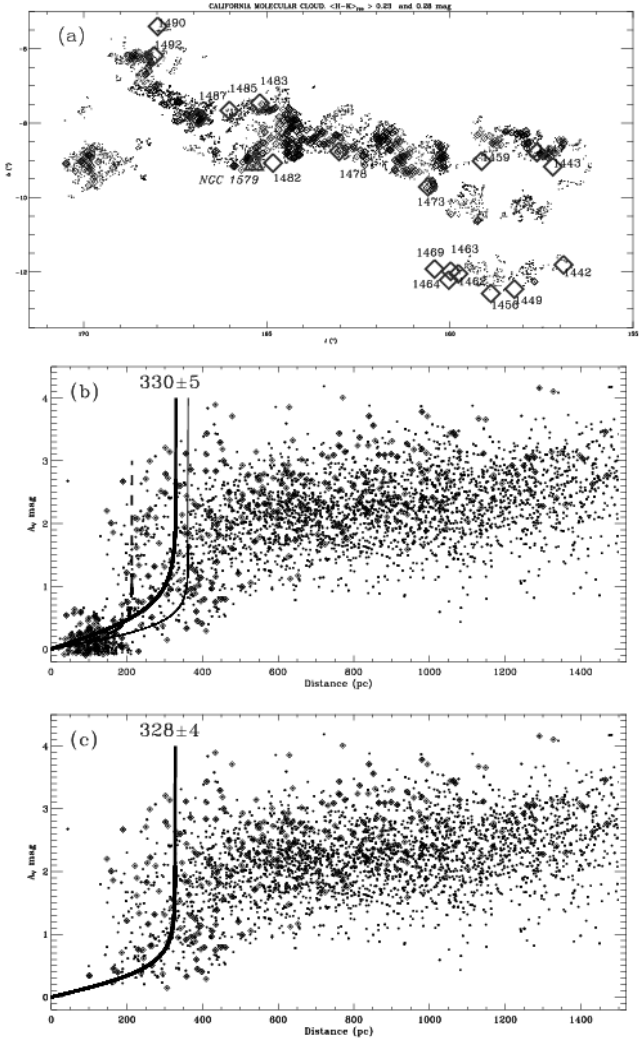} 
\caption[]{The California Molecular Cloud. (a) Collection of stars with $\sigma_{JHK}\leq$0.040 mag and 
$\overline{H-K}_{res}$ greater than 0.23. Small diamonds are stars located in 
reseaus with $\overline{H-K}_{res}$ greater than 0.28. Large diamonds indicate
Lynds dark clouds and the triangle marks the position of NGC 1579.
(b) Resulting extinction -- distance pairs from the $\overline{H-K}_{res}$ $>$ 
0.23 mag, $\sigma_{JHK}$ $<$ 0.040 sample. $D_{max}$ = 500 implies $D_{CMC}$ = 
330$\pm$5 pc (thick solid curve). The dashed curve corresponds to 
$D_{NGC 1333 (Perseus)}$ = 213$\pm$12 pc, see Fig.~\ref{f47}. The thin solid 
curve, $D_{CMC}$ = 362$\pm$3 pc, results when the {\it "Perseus layer stars"} 
with $D_{max}$ = 350 pc are excluded from the distance estimate.  
(c) Extinction - distance pairs for stars in reseaus with $\overline{H-K}_{res}$
$>$ 0.23 mag and with $\sigma_{JHK} <$ 0.040 mag without the M4 -- T stars. The 
solid curve indicates $D_{CMC}$ = 328$\pm$4 pc. Again the small diamonds show the
$\overline{H-K}_{res}$ greater than 0.28}
\label{f48}
\end{figure}

\section{Summary of distances to $\sim$25 local clouds}\label{summarizing}

Table~\ref{t2} summarizes distances to the clouds we have considered.
Apart from DC300.2-16.9 where a sufficient number of stars were not available
all distances  and standard deviations result from the $A_{V, \ estimate}$ = 
~$arctanh^p(D_{\star}/D_{cloud})$ fitting procedure. In the Table we have 
indicated that the Serpens cloud was used as a template for developing our 
method. 

\begin{center}
\begin{table}
\caption{Hipparcos/2mass distance estimates to nearby clouds}
\begin{tabular}{|l|l|r|}
\hline
Name&D$_{\rm CLOUD}$&$\pm$  $\sigma$$_{\rm CLOUD}$\\
    &pc&pc\\
\hline
Serpens (template)&193&$\pm$13\\
Taurus&147&$\pm$10\\
Ophiuchus&133&$\pm$6\\
LDN 204&133&$\pm$6\\
LDN 1228&235&$\pm$23\\
LDN 1622&233&$\pm$28\\
LDN 1634&266&$\pm$20\\
Lupus I&144&$\pm$11\\
Lupus II&191&$\pm$13\\
Lupus III$_{\rm A}$&205&$\pm$5\\
Lupus III$_{\rm B}$&155&$\pm$3\\
Lupus III$_{\rm C}$&230&$\pm$21\\
Lupus IV&162&$\pm$5\\
Lupus V&162&$\pm$11\\
Lupus VI&173&$\pm$10\\
Chamaeleon I&196&$\pm$13\\
Chamaeleon II&209&$\pm$18\\
Chamaeleon II&217&$\pm$12\\
Chamaeleon$_{\rm I,\ II,\ III}$&194&$\pm$9\\
DC300.2-16.9&118:: &$\pm$24 \\
Musca&171&$\pm$18\\
Southern Coalsack&174&$\pm$4\\
Circinus&436&$\pm$29\\
DC314.8-05.1$_{\rm 1.jump}$&372:&$\pm$52\\
DC314.8-05.1$_{\rm 2.jump}$&610:&$\pm$25\\
IC 5146$_{\rm Northern \ filament}$&603:&$\pm$68\\
IC 5146$_{\rm cluster}$&766:&$\pm$112\\
Corona Australis&148&$\pm$13\\
LDN 1459 or NGC 1333&213&$\pm$12\\
 -- part of the Perseus cloud& & \\
California Molecular Cloud&330&$\pm$5\\
\hline
\end{tabular}
\label{t2}
\end{table}
\end{center}
  
\section{acknowledgements}
This publication makes use of data products from the Two Micron All
Sky Survey, which is a joint project of the University of Massachusetts
and the Infrared Processing and Analysis Center/California Institute of
Technology, funded by the National Aeronautics and Space Administration
and the National Science Foundation. This research has made use of the SIMBAD
database, operated at CDS, Strasbourg, France.

Claus Fabricius is sincerely thanked for his contributions to the early stages 
of this work.

\begin{appendix}
\section{The ~$(J-K)_0$ $-$ ~$M_J$ ~calibration of the main sequence}
\label{appA}
\subsection{The calibration sample}
To have intrinsic colors we are obliged to use nearby, presumably unreddened, 
stars from the Hipparcos Cataloque, Perryman et al. (\cite{PMAN97}). To obtain 
a precise calibration we use stars with 
$\pi$ ~$\geq$ 0.010", assumed to imply virtually no reddening, and with
a relative error better than 10$\%$. 
The stars fulfilling these two criteria constitute
the astrometric sample. The astrometric sample is compared to the 2mass catalog 
and the common stars are extracted.

For several entries the Hipparcos Catalogue contains spectral and luminosity 
classification from the literature.  A substantial part of the stars common to 
2mass and Hipparcos does, however, not have any classification but may anyway 
be dwarfs and should be included in the sample in order to substantiate the 
main sequence calibration. Fig.~\ref{f2} shows the distribution of the stars 
without classification (gray points) overplotted the astrometric sample 
(black points). More than 5500 stars meeting the $\pi$ $\geq$ 0.010" 
and $\sigma_{\pi}$/$\pi$ $\leq$ 10.0$\%$ criteria are without a luminosity 
classification.

\begin{figure}
\epsfxsize=8.0cm
\epsfysize=8.0cm
\epsfbox{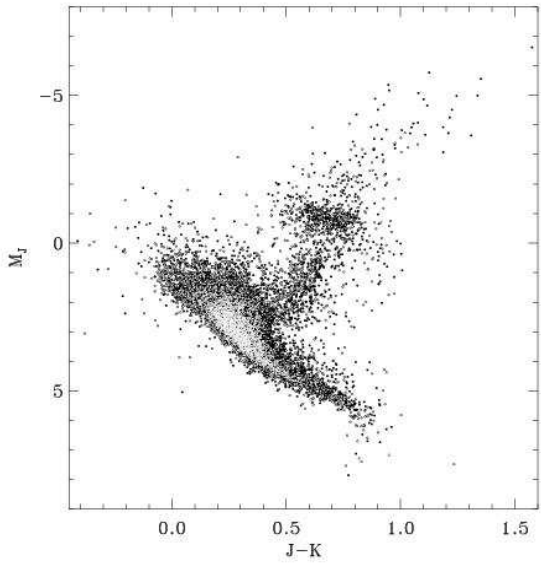} 
\caption[]{Dark points all Hipparcos stars fulfilling the astrometric criteria:
$\pi$ ~$\geq$ 0.010" and $\sigma_{\pi}$/$\pi$ $\leq$ 10.0\%. The 
overplotted lighter points are those without luminosity classification and 
many of these must be main sequence stars and should accordingly be included in 
the calibration sample}\label{f2}
\end{figure}

\subsection{Confining the main sequence}

Some stars classified as evolved are better
appreciated as main sequence stars. There is a concentration of stars classified
as LC IV as well as LC III at $(J-K, ~M_J)\approx (~0.2, ~+2)$. A possible
reason for classifying stars in the color range from about 0.1 to 0.3, the
approximate A6 -- F5 region, as "giant" like could be that they have a small
Vsini for their color? To have as many precise
main sequence stars as possible for the color -- absolute magnitude calibration
we do not rely entirely on the luminosity classification given in the Hipparcos
Catalogue but try to delineate what we think is the proper main sequence.

It is important to separate the main sequence from the subgiant
branch: the region around $(J-K, M_J) \approx (0.5, 3.0)$ 
where the MS and giant branch are separated by $\sim$3 mag in $M_J$. A mean or
median value of $M_J$ at $(J - K) \ \approx$0.5 would be located in the gap and
would represent no stars.
Stars in the Hipparcos/2mass cut are included in the calibration sample 
if located inside the main sequence demarcation as defined in the
following (result in Fig.~\ref{f8}). Apart from the apparent confinement in the 
$(J-K)_0 ~vs. ~M_J$ 
plane we corroborate "our main sequence" in two ways. It turns out that the 
$J, H$, and $K$ bands are not equally sensitive to evolution so we use the 
branching in the $M_{J} ~vs. ~M_{K}$ diagram, Fig.~\ref{f5}, for a coarse 
separation of giants from dwarfs. And we use theoretical isochrones to help 
confining the main sequence to the blue and to the bright, evolved side.

The resulting confinement is shown in Fig.~\ref{f8}. The issue for
introducing this confinement is to obtain a separation of LC V and IV in the
turn off region. The $M_J ~vs. ~M_K$ diagram 
is useful by suggesting a separation of the giant and dwarf 
sequences. In Fig.~\ref{f5} is shown how the $M_J$ and  $M_K$ magnitudes 
separate in the $M_J$ range from +3 to -2 in a giant and a dwarf branch.  
For any given $M_J$ magnitude in this range the giants are more luminous in
the K band than the dwarfs. For a distinction between the two 
luminosity classes we fitted an upper envelope to the dwarfs in 
the form of a straight line to the brightest part of the main sequence. 
According to Fig.~\ref{f2} the partition is to run in the $(J-K)$ range from 
about 0.3 to about 0.45. The lower K-luminosity limit for the giants is 
suggested as: $M_K ~\approx ~0.97575\times M_J ~-0.33265$. The solid line is the
proposed division between giants and dwarfs, the dashed line is a 45$^{\circ}$ 
line.

When the dividing line is transformed to the $(J-K)$ -- $M_J$ plane it defines 
the partition between class V and class IV. The division is shown in 
Fig.~\ref{f6}.

Stars classified as LC IV in the Hipparcos/2mass sample are overplotted as
light gray points circles in Fig.~\ref{f6} and we notice that there is a clear 
coincidence of the transformed dividing line from Fig.~\ref{f5} and the upturn 
of the subgiants indicated by a set of isochrones on Fig.~\ref{f6}.

Another important issue is how the upper luminosity limit of the early part 
of the main sequence should be identified? Important because of its influence
on the spread assigned to the estimated absolute magnitudes.  Fig.~\ref{f2} 
shows how sparsely populated it is in the Hipparcos sample. The upper 
confinement for this part of the main sequence might instead be based on all 
stars from the 2mass/Hipparcos comparison, irrespective of parallax and its 
relative error. In this sample even the blue $(J-K)$ limitation, which seems 
virtually uninfluenced by extinction is well defined. The upper bright limit 
is, however, drawn where the brightest members in the astrometric sample are 
located (see Fig.~\ref{f2} and Fig.~\ref{f6}) and not at a virtually constant 
$M_J \approx$ ~+1 where the bulk of the calibration sample has its 
bright limit.

Theoretical isochrones in the infrared, $JHK$, have been published by 
Cordier et al. (\cite{CPCS07}) and the Hipparcos/2mass main sequence
stars should be confined by a very young and moderately old isochrones. The
very young one should coincide with the youngest stars in the sample and 
delineate the lower luminosity boundary together with the blue main sequence 
confinement. The moderately old ones, younger than a few Gyr, might help locate
where A type stars and earlier types leave the main sequence. We have included 
a set of such isochrones in Fig.~\ref{f6} with ages ranging from 0.1 to 
8.0 Gyr. The 0.1 Gyr isochrone is a rather good representation of the lower 
envelope and the blue confinement. The upper confinement of the data is located
roughly where hydrogen burning in a thick shell is replaced by shell hydrogen 
burning. 
The 8 Gyr isochrone also represents the dwarf - subgiant transition rather well.
Shifting to the 12.0 Gyr isochrone does not shift this red limitation 
significantly. In the region $(J-K) \gtrsim $0.4 the isochrones follow a central
location in the MS distribution. 

Guided by these considerations we have drawn the border line
around the sample. Stars inside this curve are now considered as $the$ main
sequence sample, and will be used for the $M_{JHK} - color$ calibrations.
It is shown in Fig.~\ref{f8} together with the isochrones.
The precise location of 
the upper luminosity confinement is not that critical, except where the giants 
branch off. This is because the luminosity distribution across the main 
sequence at a given color has very few stars at the extreme luminosity. Stars
at the upper main sequence confinement evolve fast implying a low density of 
data points and they are located at the blue limit of the Hertzsprung gap.

\begin{figure}
\epsfxsize=8.0cm
\epsfysize=8.0cm
\epsfbox{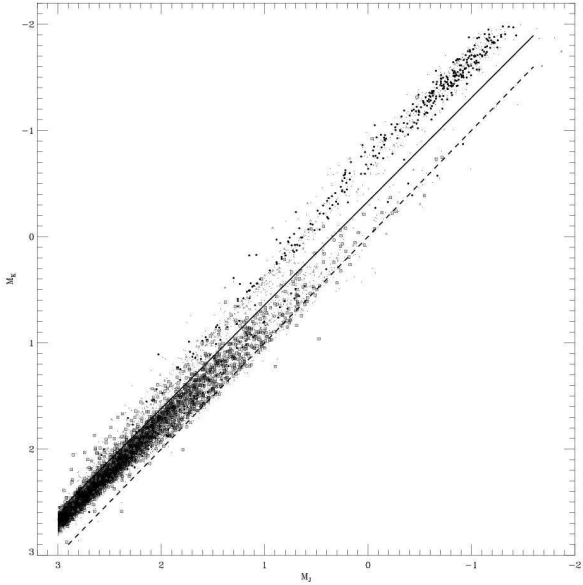} 
\caption[]{The branching part of the $M_J ~vs. ~M_K$ diagram which is
used for separating LC V (tiny boxes) from LC III (black points) 
and from LC IV (triangles). The upper solid curve is the line 
dividing dwarfs and giants and the lower dashed curve is a 
45$^{\circ}$ line.}\label{f5}
\end{figure}

\begin{figure}
\epsfxsize=8.0cm
\epsfysize=8.0cm
\epsfbox{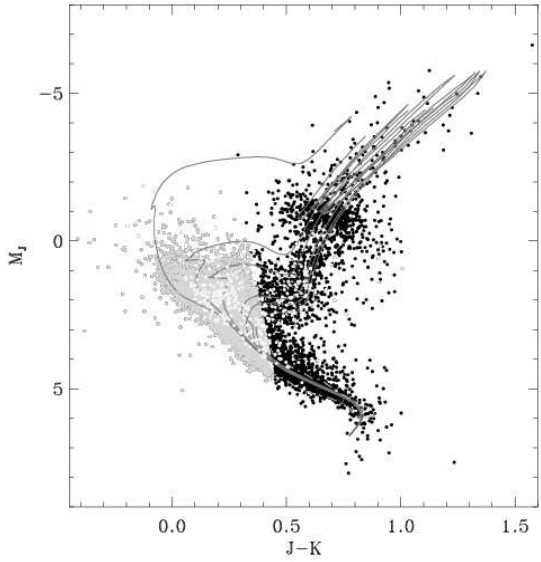} 
\caption[]{The figure shows the early main sequence resulting from the simple
dividing line in Fig.~\ref{f5}. The cool part of the main.sequence is not 
included this way because the $M_J ~vs. ~M_K$ distribution turns back to the 
high K-luminosity side of the dividing line. Luminosity class IV stars are 
overplotted as the brightest points. Isochrones of ages 0.1, 0.8, 1.5, 4.0, 5.0 
and 8.0 Gyr are from Cordier et al. (\cite{CPCS07})}\label{f6}
\end{figure}

\begin{figure}
\epsfxsize=8.0cm
\epsfysize=8.0cm
\epsfbox{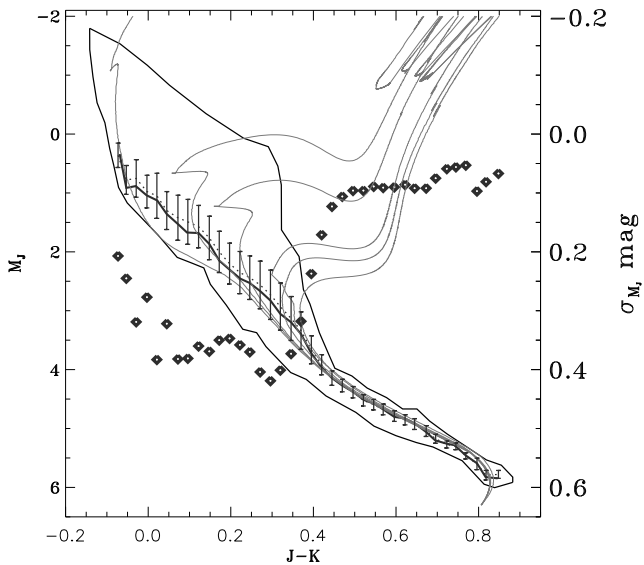} 
\caption[]{The thin solid curve is the confinement of the main sequence 
sample discussed in the text and is shown together with its resulting 
statistical relations calculated for 0.025 mag bins of $(J-K)$. The dotted 
curve is the mean given together with the standard deviation computed for 0.050 
$(J-K)$ intervals. The median curve is the solid thick one.
Isochrones from Cordier et al. (\cite{CPCS07}) are shown 
for 0.1, 0.8, 1.5, 4.0, 5.0 and 8.0 Gyr. The diamonds show $\sigma_{M_J}$ 
(right hand scale) calculated for overlapping 0.050 intervals in $(J-K)$ 
separated by 0.025 mag and a drop from 0.4 mag to 0.1 mag is noted where the 
8 Gyr isochrone turns off the main sequence}\label{f8}
\end{figure}

\begin{figure}
\epsfxsize=8.0cm
\epsfysize=8.0cm
\epsfbox{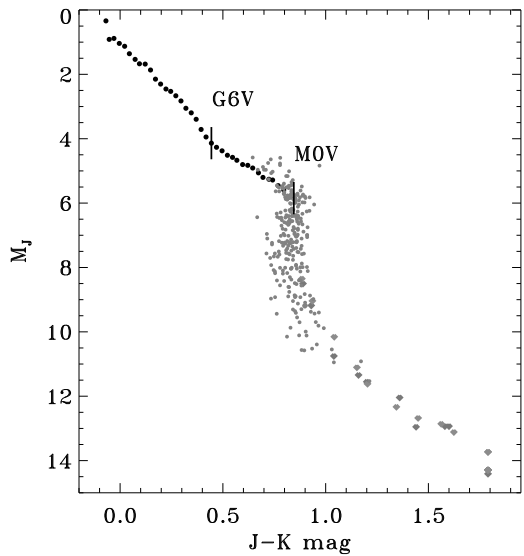}
\caption[]{For illustration we show the complete main sequence including M 
and L dwarfs. Early main sequence, black points, is from the present work. 
Grayish points are M dwarfs from Reid and Cruz (\cite{RC02}) and the cool 
tail of L dwarfs are from Leggett et al. (\cite{leggett02}) and Dahn et al. 
(\cite{dahn02}). For the L dwarfs both median and mean values are 
displayed}\label{f9} 

\end{figure}

\subsection{Mean and Median Loci}
The sample within the main sequence confinement and meeting $\pi \geq$10 mas 
and $\sigma_{\pi}/\pi\leq$0.10 criteria consists of 9085 stars. Since it is 
not possible to give a precise location of a star in the $(J-K)_0 \ vs. \ M_J$ 
diagram from only three photometric bands we replace the main sequence with a 
locus giving a representative absolute magnitude as a function of a color. Due 
to the natural width of the main sequence the replacement will introduce 
offsets from the true magnitude.
\subsubsection{Mean Locus}
After dividing the main sequence into 0.025 mag intervals in $(J-K)_0$ we 
compute the mean and standard deviation of the color and the absolute magnitude 
for each color bin. The run of the mean locus is given as the dotted curve 
in Fig.~\ref{f8}. The standard deviation is calculated for 0.050 $(J-K)_0$ 
bins though but are plotted for every 0.025 mag intervals. The standard 
deviation represents the error we commit by using the mean main sequence locus.
The standard deviation is plotted as diamonds referring to the right hand 
scale of the figure. There is a dramatic decrease in the $\sigma_{M_J}$ values 
with a factor $\gtrsim$2 in the $(J-K)$ bin 0.3 -- 0.5 from $\approx$0.4 to 
$\approx$0.1. The errors range from 
$\approx$0.06 to $\approx$0.4 for the late and early types
respectively. Paying no attention to any other error source the inaccuracy 
translates to a relative distance error $\Delta R/R$ ranging from 3\% to
20\% for individual stars with $\sigma_{J,K} <$0.035 mag. The inaccuracy will
apply to the estimate of $individual$ stellar distances. The features
whose distances interest us $-$ extinction discontinuities $-$ are defined by
several stars, maybe a number $\gtrsim$10 implying an error of the mean distance 
better than $\gtrsim$10\%. The sequence of errors narrows at about 
$(J-K)_0 \approx$0.45 corresponding to early G dwarfs. The red termination of 
the sequence is at K7 -- M0. M dwarfs are thus not included in this calibration 
but dedicated studies of their infrared absolute magnitudes have become 
avaible in the literature, Dahn et al. (\cite{dahn02}), Leggett et al. 
(\cite{leggett02}), Reid and Cruz (\cite{RC02}). Fig.~\ref{f9} shows how the 
absolute magnitudes of the M and L dwarfs fit into the present calibration. 
From the error point of view late G and K dwarfs are very well suited for the 
distance derivation but as seen in Appendix ~\ref{appB} on the $JHK$ data 
these stars may not be of immediate use due to the giant-dwarf degeneracy and 
the shape of the main sequence in the two color $(H-K)$ ~-- ~$(J-H)$ diagram. 
In Sect.~\ref{appB3} and \ref{appB4} we
suggest how these stars possibly may be included in the distance derivation. 

\subsubsection{Median Locus}
With the same color binning as for the mean we calculate the median color 
and absolute magnitude for each bin. The solid black curve of Fig.~\ref{f8} 
shows the resulting median curve and for the early part of the diagram we 
notice that the median locus is slightly fainter than the mean which was to be 
expected. 

\subsection{Dispersion in the distance calibration}

As a test we have applied the median calibration on the calibration
sample itself following the prescription of Subsection \ref{AnAlgo}. Only for 
spectral types earlier than $\sim$G6V though, by running the sequence of codes 
used for the distance and extinction estimates we have developed for the 2mass 
data. Note that this spectral range has the most imprecise calibration with
$\sigma_{M_J}$ $\approx$0.4 mag.

For this excersize we can not assume that the Hipparcos sample is unreddened 
but must estimate the intrinsic colors, $(H-K)_0$ and $(J-H)_0$, as we do with 
any
2mass extraction. We have no demands to the accuracy of the $JHK$ photometry.
From the $(J-K)_0$ vs. $M_J$ calibration we have the distance estimate which we 
compare to the trigonometric distance $\pi^{-1}$. The mean difference of 
these distances becomes 8.8 pc and the standard deviation about 25 pc. 
The dispersion of the mean differences is $\lesssim$10 pc and derives almost 
exclusively from the astrometric errors. Since the mean difference only differs
from zero on the sigma level we have not decreased the calibrated distances 
with the zero point offset. If we subtract the error coming from the 
trigonometry the 
dispersion of the distribution of residuals decreases to $\approx$20 pc.
All stars being closer than 100 pc this dispersion agrees with our calculated
standard deviations of $M_J$ in the range from 0.2 to 0.4. 0.4 is the value
pertaining to stars earlier than G6V. $\sigma_{M_J}$=0.4 implies a 
relative uncertainty in a single distance of 18$\%$.

\section{The $JHK$ data}\label{appB}

Knude and Fabricius
(\cite{KF03}) presented a preliminary discussion of the Hipparcos/2mass 
combination applied to distances of interstellar features. 
For the $JHK$ extraction an oversized area in the direction under 
investigation is defined and the errors and flags to be accepted are selected. 
For clouds in the solar vicinity outlines 
are known from mm observations of $^{12}CO$ rotational lines or from optical or
infrared extinction maps. If not available the complete set of {\it JHK} 
observations itself offers an estimate of the outline either from simple star
counts or from contours of mean values of $(H-K)$ formed in a reseau centered 
on each extracted 2mass star.

It has proven to be of some importance leaving out the photometry with the 
largest errors for the distance and extinction derivation
whereas the complete sample is retained for the star counts.
Most often we base $\overline{H-K}_{res}$ contours on stars with
precise photometry $\sigma_{JHK} \leq$0.04 but the limit is sometimes relaxed to
0.06 or even to 0.08 to include a larger number of stars.

Star counts are done in circular reseaus required to contain  
100 counts on the average. A count is assigned to each entry in the extraction 
from the 2mass catalog thus leaving out a margin the size of the radius of the 
reseau of the originally defined area on the sky. The reseau radius is 
typically $\lesssim$10$\prime$. The reseau size depends on the galactic 
latitude ranging from $\sim$15$\prime$ at the poles
to $\sim$5$\prime$ close to the plane. Even a change of a few degrees in 
latitude may change the appearence of a cloud as given by counts.
Neighboring counts are thus not independent but the stars outlining a given 
count is only used for indicating the possible presence of extinction and not 
for evaluating its size or extinction. 

Fig.~\ref{f13} and \ref{f14} both cover two dense knots of Lupus IV located 
approxmately ($\alpha , \delta$) =(242.8, -41.7) and (242.1, -41.7). See e.g. 
Fig.~8 of Cambr{\'{e}}sy or the CO map in Fig.~2 of Tachihara et al.
(\cite{tachihara01}) or the more recent extinction map by Lombardi, Lada and
Alves (\cite{LLA08}). See also the $\overline{H-K}_{res}$ map in Fig.~\ref{f35}. 
The scales of the two Figures are identical so the suggested dimensions of the 
knots are quite different when counts less than 125/reseau is used as the defining 
limit and they are not reproduced in the $\overline{H-K}_{res}$map of the
same region, Fig.~\ref{f35}. The low declination field, Fig.~\ref{f14}, 
suggests a size four to five times larger than the high declination field, 
Fig.~\ref{f13}. This tendency may be understood as an effect of the galactic 
latitude. Increasing the
latitude will lower the average stellar density, and increase the reseau size
required to contain 100 stars on the average. 
There is no unique $a priori$ way to define the area from which stars 
may be drawn for the distance determination of the cloud. Extinction contours
drawn from mean $H-K$ reseau values may likewise be influenced by latitude
because the stellar distribution varies with latitude. A cloud confinement is 
defined by a lower $\overline{H-K}_{res}$ limit identified just outside the
cloud perimeter, see Fig.~\ref{f28}(b).  

\begin{figure}
\epsfxsize=8.0cm
\epsfysize=8.0cm
\epsfbox{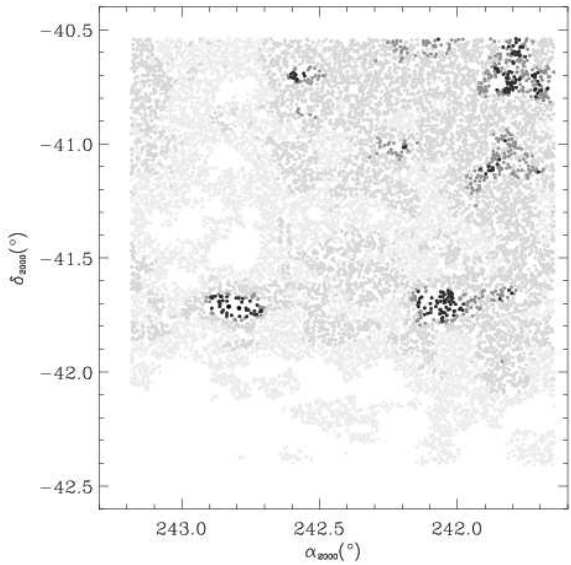} 
\caption[]{Region containing eastern part of the Lupus IV cloud. Reseaus 
displayed are those with less than 150, 125, 100 and 90 stars/reseau
respectively. Very few stars are located in reseaus with less than 75 
stars/reseau (large black dots)}\label{f13}
\end{figure}

\begin{figure}
\epsfxsize=8.0cm
\epsfysize=8.0cm
\epsfbox{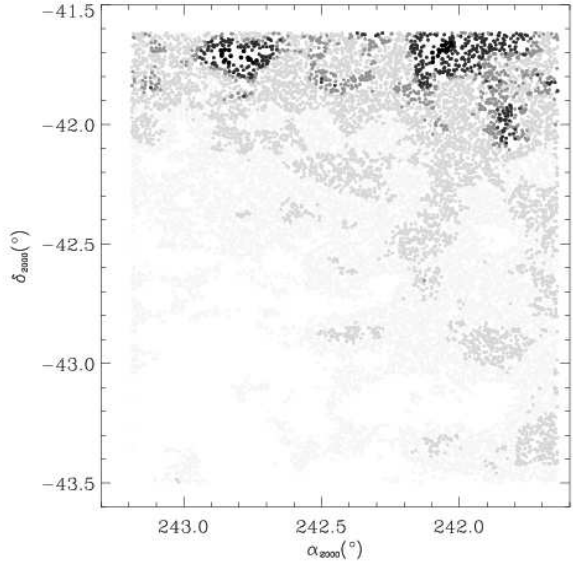}  
\caption[]{Region containing eastern part of the Lupus IV. Like the previous 
figure but shifted one degree towards the South Celestial Pole. Note that after 
the shift in declination more stars are located in reseaus with less than 75 
stars/reseau (large black dots) and the associated dense clumps appear larger 
than in Fig.~\ref{f13}. $\overline{H-K}_{res}$ contours for this part of Lupus 
IV may be seen in Fig.~\ref{f35}}\label{f14} 
\end{figure}

\begin{figure}
\epsfxsize=8.0cm
\epsfysize=8.0cm
\epsfbox{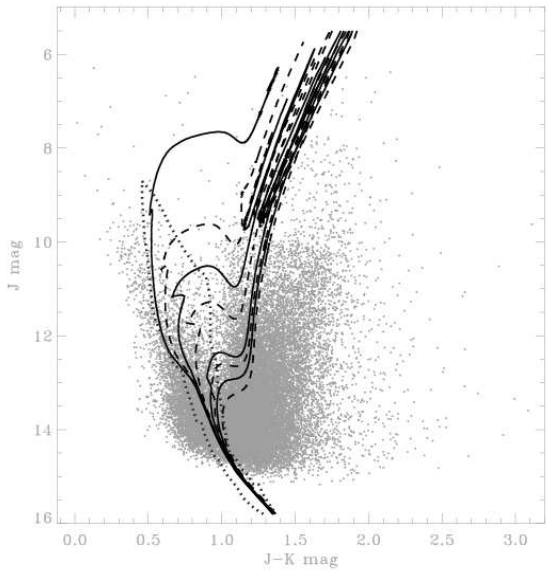} 
\caption[]{2mass data from a 2$\times$2 $\square ^{\circ}$ box
in the direction of the Serpens star forming cloud. 
($\alpha , \delta )_{center, 2000}$ = ($18^h 24^m, 0^{\circ} 0\prime)$. Slightly 
off set from the dense knots shown in Fig.~\ref{f16}. 25570 point sources with
$\sigma_{JHK} \leq$ 0.040. The main sequence confinement in $(J-K) - M_J$ are 
shown together with the 0.1, 0.4, 0.8, 1.5, 8.0 and 12 Gyr isochrones shifted 
10.5 mag in $J$ and 0.6 in $J-K$}\label{f15}
\end{figure}

\subsection{Which parts of the $(H-K)$ ~$-$ ~$(J-H)$ diagram may be used 
for the intrinsic color estimate? The Serpens region as an example}

In Fig.~\ref{f15} we have shown the color magnitude diagram $(J-K)$ vs. $J$ 
for a 2$\times$2 $\square ^{\circ}$ box confining part of the Serpens molecular cloud 
and overlaid with a set of isochrones from Cordier et al. (\cite{CPCS07}) 
for comparison. The isochrones are shifted 10.5 mag and assuming reddening of 
$E_{J-K}$ = 0.6. The 10.5 mag is chosen because with this
shift a few giants are located on the 8 -- 12 Gyr isochrone and we note that 
the shifted 8 and 12 Gyr isochrones and the upper confinement of the shifted
main sequence almost are superposed in the $(J-K)$ vs.
$J$ diagram for $(J-K)$ $\gtrsim$ 1.0. A main sequence may be identified as well 
as a very broad giant branch where the width probably is caused by extinction. 

The black points of Fig.~\ref{f17} show the $(H-K)$ vs. $(J-H)$ two color 
diagram for the same stars. The main sequence and giant 
relations from Bessell and Brett (\cite{BB88}) supplemented with the
relations from Dahn et al. (\cite{dahn02}) and Leggett et al. (\cite{leggett02})
for the cool dwarfs and from Allen (\cite{allen2000}) for the hot stars not 
contained in Bessell and Brett are overplotted. The relations have been 
transformed to the 2mass system, Carpenter (\cite{JMC01}).

Two straight lines with the slope of the reddening ratio
$E_{J-H} / E_{H-K}$=1.916 assumed to be constant for all spectral classes are 
shown. The value, 1.916, is close to the ratio derived from Fitzpatrick's
 (\cite{F99}) 
model of the extinction law computed for $R_V$=3.1 and assuming a reddening 
$E_{B-V}$=0.5. This means that we use a law pertaining to the diffuse part of
the interstellar medium.

The upper one of the two reddening vectors in Fig.~\ref{f17} 
intersects the main sequence where the giant sequence has it hottest point.
The lower one crosses the main sequence at its hottest point
and acts as a lower envelope in the color -- color diagram.

To obtain intrinsic colors a main sequence star is translated parallel
to a reddening vector to the standard locus with a shift
given by its reddening. When a star is extrapolated back to
its location on the main sequence locus its intrinsic colors $(H-K)_0$ and
$(J-H)_0$ are known. This implies that we have an estimate of the star's reddening 
in $(H-K)$ as well as in $(J-H)$. We prefer using $E_{J-H}$ which, apart
from the larger range of $(J-H)_0$ compared to $(H-K)_0$, is relatively less
influenced by the photometric errors than $E_{H-K}$. From $E_{J-H}$ and the
extinction law $A_J$ is obtained. Applying the absolute
magnitude calibration of the main sequence locus from in Appendix~\ref{appA}, 
Fig.~\ref{f8} supplemented with literature values for spectral types earlier 
than $\sim$B4 and later than $\sim$M0 we may estimate the stellar distance 
and produce a diagram showing the extinction variation with distance. 

Complications arise because the giant and main sequence relations 
overlap and because the main sequence has the shape it has, after an almost
linear dependence of $(J-K)_0$ on $(H-K)_0$ the relation breaks and $(J-H)_0$
becomes almost constant with increasing $(H-K)_0$ implying that we may not
discern a heavily reddened early main sequence star from a less reddened late
type main sequence stars. A degeneracy that will be remedied in the next
decade when trigonometric parallaxes become available for most of the 2mass
stars.

\begin{figure}
\epsfxsize=8.0cm
\epsfysize=8.0cm
\epsfbox{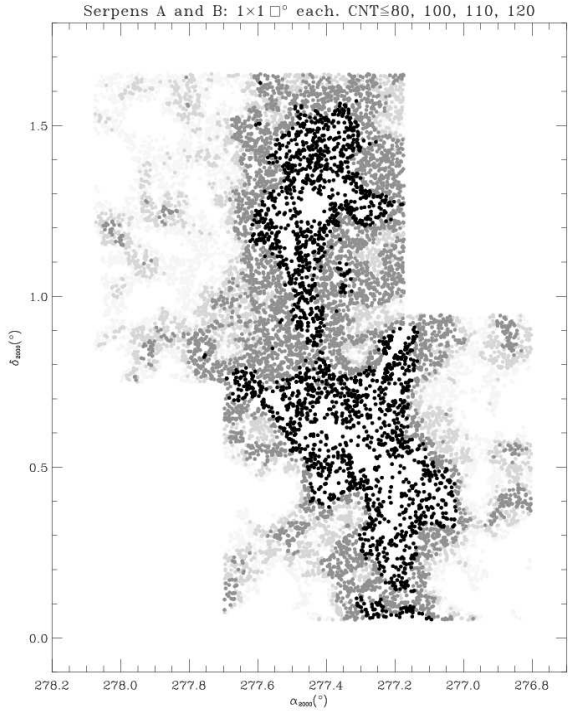} 
\caption[]{2mass data for two slightly overlapping 1$\times$1 $\square ^{\circ}$
regions in the direction of the Serpens star forming clouds A and B.
Reseaus with counts less than 80, 100, 110 and 120 are shown. The countours of 
these counts may be compared to the extinction map given by Enoch et al. 
(\cite{enoch07}) in their Fig.~5.} \label{f16}
\end{figure}

\begin{figure}
\epsfxsize=8.0cm
\epsfysize=8.0cm
\epsfbox{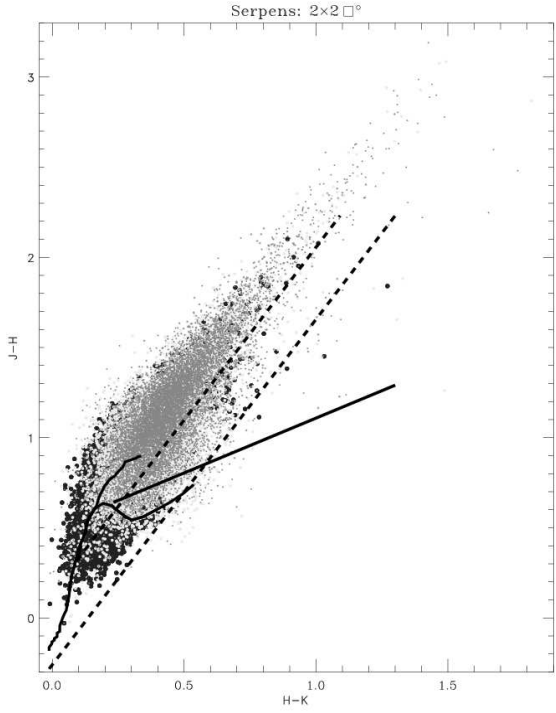} 
\caption[]{2mass data from the two 1$\times$1 $\square ^{\circ}$ boxes shown in 
Fig.~\ref{f16} in the direction of the Serpens star forming clouds: A(light gray) and B(dark gray). The main sequence and giant loci are displayed 
together with two reddening vectors. 
The upper one intersects the main sequence relation where the giant relation 
has its origin. The third line is Meyer et al.'s (\cite{MCH97}) unreddened 
T~Tauri locus. The two cloud regions are overplotted data from the nearby
2$\times$2 $\square ^{\circ}$ region shown in Fig.~\ref{f15} (black symbols) covering
the southern part of Fig.~\ref{f16}. We notice that part of the 
2$\times$2 $\square ^{\circ}$ region are much less reddened than the two 
1$\times$1 $\square ^{\circ}$ boxes}\label{f17}
\end{figure}

\subsection{The early types on the MS}

To avoid the mismatch of giants and dwarfs only stars below the upper reddening
vector of Fig.~\ref{f17} should be applied. And to avoid the early/late type 
dwarf mixing
we should limit the study to reddenings pertaining to the early dwarfs located 
between the two parts
of the main sequence locus. This implies a bias in the reddening range that may 
be probed. The hottest part of the main sequence may trace reddenings equivalent
to $A_V \lesssim$ 6 mag whereas the dwarfs located just where the giant relation
branches off only measure $A_V \lesssim $ 3 mag but these values are sufficient
to identify an extinction discontinuity. We impose a mininum distance
measured along a reddening vector from the cool part of the relation in order to
assure that a point is not caused by the error distribution among the late M 
dwarf. Similarly we introduce a minimum distance in $(J-H)$ from the upper 
reddening line. We also exclude stars located to the blue side of the main 
sequence in terms of the $(H-K)$ color. The sample located between the two 
reddening vectors of Fig.~\ref{f17} and confined by the main sequence
locus and reduced by the imposed margins is our prime tracers of reddening and
distance. These stars do, however, belong to the less populated bright part of 
the luminosity function with a rather
low spatial density. They may trace rather large volumes and measure
substantial extinction values but for nearby, small molecular clouds typically
with solid angles smaller than a few square degrees, the
probability to find such stars in front of the cloud is small. And
for a good estimate of a cloud distance unreddened stars in front and reddened
stars just behind the cloud are required. Stars with a larger spatial density
may be required to trace the volume in front of any cloud. Experiments on most 
of the local star forming clouds have shown that the O -- G6 MS range most often
do not provide enough stars in front of the extinction jump. This means that
the cooler K and M dwarfs should be considered in order to provide an estimate 
of a lower distance estimate to the cloud.

\begin{figure}
\epsfxsize=9.0cm
\epsfysize=6.0cm
\epsfbox{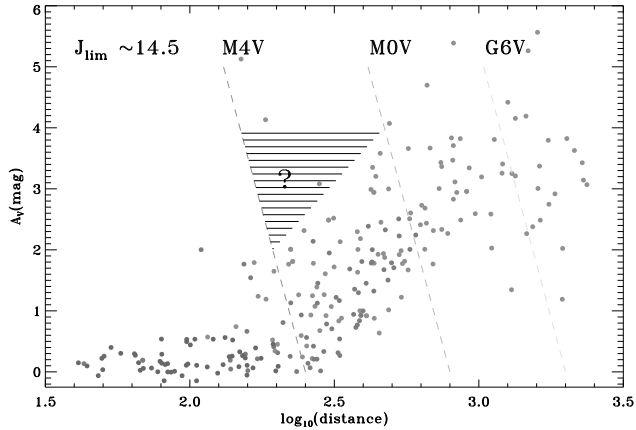} 
\caption[]{Extinction vs. distance data combined for the two regions Serpens
cloud A and B shown in Figs.~\ref{f16} and \ref{f17}. The three dashed lines 
show the maximum extinction observable for three divisions in the 
$(H-K) - (J-H)$ diagram. The central line is caused
by the turn over of the $(H-K) - (J-H)$ relation at the maximum main sequence 
$(J-H)$ value roughly corresponding to M0V. The right most corresponds to the
spectral type where the the giant locus and the main sequence coincide, 
approximately at G6V, and finally the left most 
displays the maximum extinction measurable with main sequence tracers of 
type M4V. The upper extinction limits are valid for a limiting magnitude 
$J_{lim} \sim 14.5$. The hatched triangle may only be measured with stars earlier
than $\sim$M4V, see discussion in the text}\label{f18}
\end{figure}

\subsection{The cool dwarfs} \label{appB3}
Due to the scarcity of the early type dwarfs, few are expected in the volume in 
front of a cloud implying that only upper distance limits can be provided for
most local molecular clouds. We therefore may try to include the M dwarfs as 
well. Intrinsic color and absolute magnitude are derived as for the early type 
dwarfs, the only difference is that we have replaced the independent parametera
 $(H-K)$ with $(J-H)$ because of the near $(J-H)_0$-constancy of these cool 
stars. We thus include stars with $(J-H)$ located above the main sequence value
reduced by 
$\Delta(J-H)$ = 0.070 for a given $(H-K)$ value and 0.040 mag below the upper
reddening vector in Fig.~\ref{f17}. Stars below the main sequence by more than 
$\Delta(J-H)$ = 0.070 are included in the sample of main sequence stars earlier
than G6V. The chosen limitations depend on the maximum photometric errors. As a 
reference we use $\sigma_{JHK,max}$ = 0.040 mag.

Since the volume of the molecular cloud, in the
solid angle we study -- which often is only a few square degrees 
-- typically is small, the volume in front of the cloud may be used as an
approximation to this volume plus the cloud volume and we may compare the 
number of early dwarfs (hotter than $\sim$G6) to the number of late dwarfs 
(cooler than K9/M0) from the local luminosity function. The late/early 
ratio becomes $\approx$8. Noting that the early ones that are mistaken for late
ones are the ones that have their $A_V$ exceeding the range from 3 to about 6 
mag and are emerged in the cloud with a volume that is much smaller than the 
volume in front of the cloud or are located behind the cloud so the late/early 
ratio is in fact larger than 8. The stars clustering around the lower reddening
vector of Fig.~\ref{f17} illustrate that contamination of the cool sample by 
very extincted early type stars, $A_V \lesssim$15 mag, takes place. Another 
interpretation is that these are not early dwarfs but rather on the AGB, 
Lombardi, Lada and Alves (\cite{LLA2010}). 

Since some molecular clouds are star forming, another,
less serious ambiguity arises from the presence of PMS stars. Stars with 
$(J-H)$ redder than indicated by the T Tau locus given by Meyer, Calvet and 
Hillenbrand (\cite{MCH97}) are consequently excluded. So only the stars located
between the locus defined by the main sequence relation corrected
with $\Delta(J-H)$ = 0.070 and the T Tau line is considered for the 
distance -- extinction determination. This of course biases the M dwarfs 
included in the sample to the lesser reddened ones. A nice example showing 
some very local unreddened M dwarf candidates for a Serpens region is shown in 
Fig.~\ref{f18}. We call these secondary tracers the M4 - T sample. 

The two samples we have considered so far are firstly the one constituted by 
the stars confined by the main sequence and the two reddening vectors 
originating where the giant relation branches off the main sequence and where
the hottest star included in the main sequence relation are located: this is our
primary sample for which we may estimate extinction and distance in a unique 
way. Secondly we include the dwarfs later than $\sim$M4 and confined by the main
sequence offset by -0.070 in $(J-H)$ and by the T Tau sequence. This sample may 
not be clean since it may be contaminated by dwarfs earlier than $\sim$G6 and 
with an extinction exceeding $\approx$3 -- 6 mag in V depending on the spectral
class. The contamination may not bias the results seriously since assuming 
that all stars are $\sim$M dwarfs only will imply
a wrong type in a few cases, about one out of eight.
Mistaking an O -- G6 dwarf for an M dwarf will in fact not influence the
location of the extinction discontinuity, e.g. the one in Fig.~\ref{f18}, it 
will replace a large extinction (the true one) with a small (a false one) but 
put it at a false small distance, due to the intrinsic faintness of the $M$ 
dwarfs (Fig.~\ref{f9}), where it will not influence the estimate of the cloud 
distance seriously. 

\subsection{Indications from the K dwarfs}\label{appB4}
Leaving out a spectral range in a magnitude limited sample, as 2mass, may
introduce selection effects influencing the distance -- extinction variation 
we are looking
for. Furthermore the G6 -- MO dwarfs have a high spatial density and if they
could be included in the distance -- extinction determination they would 
substantiate the presence of any extinction discontinuity suggested by the 
O -- G6V and M4 -- T samples. Not to mentioned that G6 -- M0 range has a more
precose calibration than the O -- G6 range.   

As we can see from the $(J-K)_0$ vs. $M_J$ relations in Fig.~\ref{f9} 
there is a homogeneous variation of $M_J$ with $(J-K)_0$ also in the spectral 
range from G6 to M0. This means that the exclusion of the $\sim$G6 $-$ M0 part 
of the main sequence will introduce a gap in the distance
distribution of the extinction tracers because stars with $M_J$ between 4 and 6
are systematically missing. There may, however, be a way that the K
dwarfs can be used to corroborate the distance -- extinction indications 
suggested by the early and the late steller types: If the distance vs. $A_V$ 
diagram based on stars earlier than $\sim$G6 and later than $\sim$M4 indicates
a well defined extinction discontinuity, refer to Fig.~\ref{f18} and
19 where we notice the dwarfs later than $\sim$M4 within $\sim$200 pc and the
O -- G6 dwarfs beyond $\sim$200 pc, we have an indication of the distance 
range over which the extinctions in the jump are measured. 

In order to make use of the $\sim$G6 -- M0 dwarfs we first extract the stars 
between the upper reddening line and the main sequence, see Fig.~\ref{f17}. 
This extraction is of course a mixture of dwarfs and giants. If on the
main sequence they are of spectral types $\sim$G6 $-$ $\sim$M0. 

With the distance -- extinction information deduced from the O -- G6 and 
M4 -- T samples we have an indication of how extinction varies with distance. 
We know part of the extinction range within a given distance interval. In 
Fig.~\ref{f19} we see that $A_V$ is increasing from $\approx$0.3 to 
$\approx$3.5 
mag within the distance interval from $\sim$60 pc to $\sim$450 pc. Given the 
$A_V$ and distance limitations we ask if any of the stars we just extracted
between the upper reddening vector and the main sequence can be located in this
distance interval (60 -- 450 in our example) and with an extinction in the range
suggested by the M4 -- T and O -- G6 dwarfs respectively.
Assuming they are dwarfs, i.e. that they obey the $(J-K)_0$ -- $M_J$ calibration
valid for the G6V -- M0V stars, we extrapolate back to the main sequence
standard curve and obtain an estimate of the intrinsic colors and subsequently
absolute magnitude just as was done for the O -- G6 range. Extinctions estimated
for the K dwarfs this way are limited to the range from $\sim$0 to $\sim$3. 
The spatial density of the $\sim$K
dwarfs are approximately the same as that of the early group. By imposing the 
distance and extinction limits in the extraction of possible K dwarfs we may
exclude the giants. The $(J-H)$ color range of these stars include the 
G0III -- K2/3III spectral range for the giants implying that the ratio of the
number densities of the G6V -- M0V to the number density G0III -- K2/3III stars
is $\approx$36.5. The LC III stars are of course brighter than LC V stars, for
$(J-K)\approx$0.7 the red clump giants are $\approx$ six magnitudes brighter
than the corresponding point on the main sequence. This means that since the
observed $J$ magnitude does not depend on whether the target is a dwarf or a
giant, since the extinctions for the observed $(H-K, J-H)$ depend little on 
whether the source is a dwarf or a giant (recall that the coincidence of the 
dwarf and the giant loci is what causes our actual problem) the relative 
distance is only determined by the difference in absolute magnitude meaning 
that if the star is a giant it is a factor $\approx$16 farther away than if it 
was a dwarf. Due to the ratio of the number densities the volume within the 
cloud distance based on the O -- G6V stars will only suffer a slight 
contamination. Typically less than 5 contaminants are expected in front of a 
cloud at 150 pc and projected into one square degree on the
sky. The volume beyond the cloud may contain a larger number of giants but here
the extinction may work in our favour. For a cloud at a distance 
$\approx$200 pc the giants may sample a volume out to about 2.5 kpc meaning 
that they may pick up an additional extinction of $A_V$ = 2 -- 3 mag from the 
diffuse medium alone. This additional extinction will may push them beyond the 
upper extinction limit of $A_V\approx$3 mag pertinent to the zone between the 
upper reddening vector and the main sequence locus so we are pretty confident 
that most of the K dwarf candidates are real dwarfs.   

An example of the resulting (1/$\pi_{JHK}, ~A_V)$ distribution from the three
groups of tracers is shown in Fig.~\ref{f19} for the 2$\times$2 $\square ^{\circ}$ area
centered on $(\alpha ,~\delta)_{center} = (18^h 24^m ,~0^{\circ} 0\prime)$ 
adjacent to clouds A and B, Harvey et al. (\cite{harvey07}) in the star forming
Serpens cloud. The color -- color diagram for 2$\times$2 $\square ^{\circ}$ is the 
black underlying points in Fig.~\ref{f17}. We note that the extinctions are 
different in these three Serpens regions. The regions containg the A and B 
clouds are generally more obscured than the  2$\times$2 $\square ^{\circ}$ area. 

\begin{figure}
\epsfxsize=8.0cm
\epsfysize=8.0cm
\epsfbox{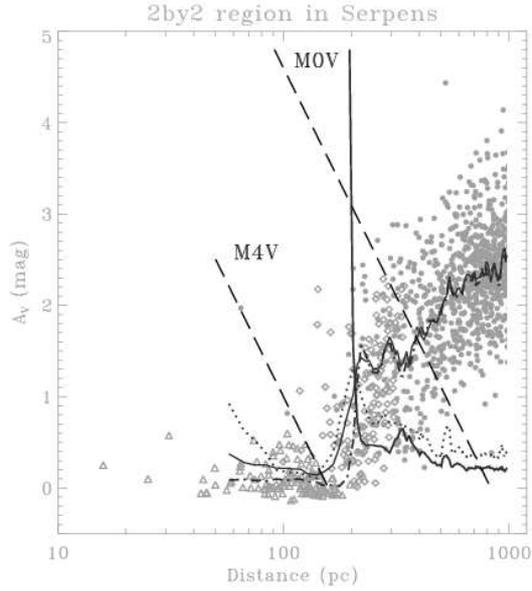} 
\caption[]{Distance vs. extinction diagram for the 
2$\times$2 $\square ^{\circ}$ region centered on part of the star forming Serpens
cloud with $(\alpha ,~\delta)_{center, 2000} = (18^h 24^m ,~0^{\circ} 0\prime)$. 
The two-color diagram for these stars region may be seen in Fig.~\ref{f17}.
Triangles are the $M4-T$ dwarfs, diamonds the proposed G6 -- K9/M0 dwarfs and
finally the larger black filled circles are the dwarfs 
earlier than G6. The distance axis is shown on a logarithmic scale in order to 
emphasize the closest stars. The two dashed lines display the maximum 
extinction observable with a $M4V$ and a $M0V$ tracer respectively with 
$J_{lim}$ = 14.5 mag. The statistics are sampled in 20 pc distance bins with 
distance steps of 10 pc. Each distance bin has a 50$\%$ distance overlap 
with each of its adjacent bins. The solid curve displays the run of the 
mean extinction $<A_V>$, the dashed curve is the variation of the median 
extinction. The dotted curve is the standard deviation/sqrt(N-1) scaled with a
factor five for clarity. Finally $\sigma_{A_V}/ \overline{A_V}$ is shown as the solid 
curve in black. The statistics for this curve are clipped below 100 pc where 
it shows some oscillations} \label{f19}
\end{figure}
\end{appendix}
\end{document}